# Systematic Coarse-graining of Epoxy Resins with Machine Learning-Informed Energy Renormalization

Andrea Giuntoli[1,2], Nitin K. Hansoge[2,3], Anton van Beek[2,3], Zhaoxu Meng[4*], Wei Chen[2,3*], Sinan Keten[1,2,3*]

[1]Dept. of Civil & Environmental Engineering, Northwestern University, 2145 Sheridan Road, Evanston, IL 60208-3109

[2]Center for Hierarchical Materials Design, Northwestern University, 2205 Tech Drive, Evanston, IL 60208-3109

[3]Dept. of Mechanical Engineering, Northwestern University, 2145 Sheridan Road, Evanston, IL 60208-3109

[4]Dept of. Mechanical Engineering, Clemson University, 208 Fluor Daniel EIB, Clemson, SC 29634-0921*To whom correspondence should be addressed. Emails:

zmeng@clemson.edu (Z. Meng)

weichen@northwestern.edu (W. Chen)

s-keten@northwestern.edu (S. Keten)

# ABSTRACT

A persistent challenge in molecular modeling of thermoset polymers is capturing the effects of chemical composition and degree of crosslinking (DC) on dynamical and mechanical properties with high computational efficiency. We established a coarse-graining (CG) approach combining the energy renormalization method with Gaussian process surrogate models of molecular dynamics simulations. This allows a machine-learning informed functional calibration of DC-dependent CG force field parameters. Taking versatile epoxy resins consisting of Bisphenol A diglycidyl ether combined with curing agent of either 4,4-Diaminodicyclohexylmethane or polyoxypropylene diamines, we demonstrated excellent agreement between all-atom and CG predictions for density, Debye-Waller factor, Young's modulus and yield stress at any DC. We further introduced a surrogate model-enabled simplification of the functional forms of 14 non-bonded calibration parameters by quantifying the uncertainty of a candidate set of calibration functions. The framework established provides an efficient methodology for chemistry-specific, large-scale investigations of the dynamics and mechanics of epoxy resins.

**KEYWORDS:** energy-renormalization, coarse-graining, epoxy, machine learning, functional calibration, mechanical properties.

# INTRODUCTION

Computational design of high-performance epoxy resins calls for methods to circumvent costly experiments. Chemistry-specific molecular models are critically needed to bridge the gap in scales between molecular dynamics (MD) simulations and experiments, while predicting accurately the highly tunable macroscopic properties of epoxy resins and their composites[1-3]. This remains a challenging problem to tackle due to the chemical complexity[4-6] of epoxy resins, the high number of properties that must be targeted for realistic predictions, and their strong dependence on the degree of crosslinking (DC) [7-12]. This up-scaling problem requires multi-dimensional functional calibration, taking inputs from high-fidelity simulations such as all-atomistic simulations. All-atom (AA) MD simulations have demonstrated great success in predicting the effect of DC on the glass transition temperature ($T_g$), thermal expansion coefficient and elastic response[13,14] of epoxy resins, and the fracture behavior of epoxy composites[15,16]. This makes AA-MD suitable for informing larger-scale models, provided that the data required for upscaling is not prohibitively expensive to obtain. While theoretical tools such as time-temperature superposition have been instrumental in bridging temporal scales [17,18], AA simulations on their own remain prohibitively expensive for high-throughput design.

Systematically coarse-grained (CG) models can extend the length and time scales of MD simulations by orders of magnitude, but chemistry-specificity requires calibration of a complex force-field to match the properties of underlying AA simulations or experimental data. Most CG models proposed for epoxies matched the structural features[19] or the thermomechanical properties [20,21] for highly-crosslinked networks. Prior models have generally not addressed the question of transferability of the model over different temperatures or curing states, which is challenging because of the smoother energy landscape and reduced degrees of freedom of CG models compared to AA models[22,23]. This particular aspect requires a functional calibration of the force-field parameters against DC, temperature ($T$), or any other variable over which transferability is desired. Machine Learning (ML) tools can efficiently handle such a parametric functional calibration in a complex force field. Despite the growing interest in utilizing ML approaches to CG modeling[24-26], complex chemistries such as epoxy resins have not been explored extensively. Progress was made on this issue in a recent epoxy CG model[27] where a particle swarm optimization algorithm was used to calibrate a $T$-dependent force-field for three different curing states with elastic modulus as the only target property. A general CG framework for epoxy resins that can target multiple properties at different DCs and demonstrate the method for more than one cure chemistry remains to be established. An accurate description of the dynamics and mechanical properties of partially cured epoxies is particularly relevant in the context of epoxy-based composites, where the exploitation of partial and multi-step curing processes can lead to enhanced performance of the epoxy resin for storage, additive manufacturing or functionalization[28]. Additionally, a model that can account for differences in curing degree across the material can be used to capture gradient properties within interphase regions of composites like CFRP[29].

To address this issue, here we simultaneously target the DC-dependence of density, dynamics, modulus, and yield strength of two model epoxy resins. A parametric functional calibration requires the functional form to be defined a priori[30,31]. This is not required by non-parametric methods that construct the calibration functions through a reproducing kernel Hilbert space[32,33]. However, either approach requires additional assumptions when used to calibrate functions in high-dimensional spaces to avoid identifiability issues[34-36]. For this reason, we employ a physics-informed strategy, leveraging our recently developed energy renormalization (ER)[37] method, which calibrates the non-bonded interactions of the CG model in a $T$-dependent fashion to match the underlying AA simulation. Based on the generalized entropy theory of the glass formation[38,39], the variation of the cohesive interaction of the CG model with varying external parameters allows to tune the activation energy of the system, which compensates for the different entropic variations of the AA and CG models caused by the different resolution of the energy landscape.

Recent ER models for different homopolymers[40], molecular glass-formers[41], and biomimetic copolymers[42] matched the mean square displacement at the picosecond time scale, $\langle u^2 \rangle$, to also predict dynamical and mechanical properties. This is because $\langle u^2 \rangle$ is strongly connected to diffusion[41], relaxation time[43-45], shear modulus[37], and vibrational modes[46] in glass-formers.

Here we extended the ER protocol to a CG model for epoxy resins, focusing on the DC-transferability and simultaneously matching the density, dynamics, and mechanical properties of the systems. We supported our protocol with the use of Gaussian processes for the calibration of the force field. This particular ML technique is extremely efficient in treating high-dimensional parametrizations, and naturally incorporates multi-response calibrations. Details of our protocol are reported in the Methods section. We targeted a system with Bisphenol A diglycidyl ether (DGEBA) as the epoxy and either 4,4-Diaminodicyclohexylmethane (PACM) or polyoxypropylene diamines (Jeffamine D400) as the curing agents. We focused on this versatile system because recent experiments[47-49] on resins prepared using a combination of PACM and Jeffamines of varying molecular weight showed remarkable mechanical properties stemming for dynamical heterogeneities at molecular scales not easily accessible to AA models. For the DC-dependent parameters of the CG force field, we initially assumed a relatively high dimension and flexible class of radial basis functions. For uncertainty quantification purposes, we calculated the fluctuations of the Gaussian process prediction in response to perturbations of the optimal solution.

This information was then used to simplify calibration functions while maintaining a comparable degree of accuracy.

The manuscript is laid out as follows. We first report the target properties from AA simulations at different values of DC from 0% to 95%. Then, we define the parametric range for the non-bonded parameters of the CG models and determine the sensitivity of the target properties on the CG parameters in this 15-dimensional range. We train surrogate ML models based on the CG and AA simulations and we report the optimal functions for all the non-bonded parameters. Using uncertainty quantification, we simplify the functional form of the parametrization, resulting in only 21 free parameters needed to calibrate 14 functions. We show that the optimized CG model has excellent agreement with all eight (8) target macroscopic properties from the AA simulations. Finally, we also show that optimal parameters for the target properties also provide a reasonably good match between AA and CG curves for the complete mean square displacements and stress-strain response datasets.

**RESULTS**

**All-atom model target properties**

The CG model for the proposed double curing agent epoxy resin system contains 7 types of beads, and 7 types of bonds and 10 types of angles among them. We aim to functionally calibrate the parameters of the CG model to simultaneously capture the DC-dependent density, $\langle u^2 \rangle$, Young's modulus, and yield stress at $T$=300K of an underlying AA model. The AA force field here employed has been validated for similar epoxy systems[50], showing that it captures the glass transition temperature and fracture behavior of experimental systems. Details on the AA model are given in our Methods section. We first calibrate the bonded parameters using a standard Boltzmann inversion (BI) approach. More importantly, the non-bonded parameters calibration was done using Machine-Learning (ML) Gaussian process models as they are data-efficient[51,52] and enable the quantification of the modeling uncertainties intrinsic to MD simulations[53]. To manage the high dimensionality of inverse functional calibration, we employ a statistical inference approach to simplify the underlying function forms. We report a scheme of our CG model and a flowchart of our parametrization process in Figure 1.

The first step in the calibration of the CG force field was to set the parameters of the bonded potentials, which was done through a BI[54] approach, to match the probability distributions informed from AA simulations. The details of the bonded terms parametrization are fully reported in our Supplementary Note 1 and Supplementary Figure 1, and the potential form and parameters are listed in Table 1.

To determine the non-bonded parameters, we first extracted initial values for the cohesive energies and bead sizes $[\varepsilon_i, \sigma_i]$, $(i = 1, ... , 7)$ from the AA radial distribution functions of all seven CG beads of the model using BI. These non-bonded parameters correctly reproduce the structure of the AA system in CG representation but fail to capture the macroscopic dynamics and mechanical properties of the system. This inadequacy makes the model insufficient to extract quantitative information from the simulations and guide the experimental design of these materials. In this study, we treat the non-bonded force field parametrization as a multi-objective optimization problem where we aim to determine 14 parameters $[\varepsilon_i, \sigma_i]$, $(i = 1, ... , 7)$ to simultaneously match the target density, Debye-Waller factor $\langle u^2 \rangle$, Young's modulus, and yield stress at all DCs.

Figure 2 reports the values of density, $\langle u^2 \rangle$, Young's modulus, and yield stress of the AA systems for DGEBA+PACM and DGEBA+D400. We note that the values found for the Young' modulus of the high DC systems are in line with experimental results[47,49], in the range of 2.5 to 3 GPa. For both systems, the density and mechanical properties increase with increasing DC, while $\langle u^2 \rangle$, a marker of mobility, decreases. This is expected, and more pronounced in the DGEBA+PACM system, which has stiffer and less mobile chain networks due to the rigidity of the curing agent PACM. Flexibility introduced by D400 increases mobility and reduces density as well as mechanical properties of the DGEBA+D400 system[47]. A quantitative comparison of $\langle u^2 \rangle$ between simulations and future experiments should be done with caution, since in experiments $\langle u^2 \rangle$ is extracted from the neutron scattering intensity[55], can depend on the scattering wavelength $Q$ and the very definition of Debye-Waller Factor includes the whole exponential term $\text{DWF} = \exp(-\frac{Q^2 \langle u^2 \rangle}{3})$, while it is customary for molecular simulation studies to use the term DWF as a definition of the $\langle u^2 \rangle$ value extracted from MSD functions[56].

Young's modulus in particular changes differently depending on DC in the two systems, since the spatial density of crosslinks is higher in the DGEBA+PACM system due to the lower molecular weight of PACM compared to D400. In other words, because of the different chain configurations of the curing agent, increasing DC leads to different changes in configurational entropy caused by the reduction in degrees of freedom. In addition, we observe that the dependence of Young's modulus on DC is nonlinear, indicating complex changes of configurational entropy with increasing DC in the epoxy resin networks.

**Non-bonded CG force field: sensitivity analysis**

Any fixed parametrization of the CG model is not able to match the properties of the AA system at all DC values, as we show in Supplementary Figures 2 and 3 in our Supplementary Note 2. This is arguably caused by the different rate with which the configurational entropy of the AA and CG models changes with varying DC, similarly to what happens with varying $T$[37]. Thus, we introduced a DC-dependence for all non-bonded parameters $[\varepsilon_i, \sigma_i] = [\varepsilon_i(\text{DC}), \sigma_i(\text{DC})], (i = 1, ..., 7)$. In previous models with highly homogeneous polymers and few CG bead types, it was possible to study the dependence on temperature with manual parameter sweeps. ER in these circumstances required only one $T$-dependent function to rescale all cohesive energies (the $\varepsilon_i$) and another to rescale all the effective sizes of the CG beads (the $\sigma_i$). We found that this was not possible in our current epoxy model due to the high complexity of the system, including the effect of crosslinks and the large amount of CG beads with different cohesive energies and sizes. Here, we introduced a generalization of previous protocols that relies on ML to explore the high-dimensional space of the model parameters. The idea is to surrogate the AA and CG models with Gaussian random processes followed by minimizing the difference between the CG and the AA models for all DC with respect to the calibration functions. Preserving the seminal idea of the ER procedure, the protocol outlined in this paper can be easily generalized to any CG model. We used the simulation data presented in Figure 2 to train the AA Gaussian process models: 19 samples for the DGEBA+PACM system and 20 samples for the DGEBA+D400 system. In the AA model, DC is the only input variable. For the CG model, DC and the non-bonded parameters $[\varepsilon_i, \sigma_i]$ are the input parameters. The range of the parameters was determined by preliminary simulations calibrating the cohesive energies either to match the dynamics of the AA systems at DC=0% or the Young's modulus at DC=90% or 95% (the highest DC we can achieve for the DGEBA+PACM or DGEBA+D400 AA networks respectively). This gave us extremes for the values of cohesive energies $\varepsilon_i$, and we further expanded them by around 20%. We also selected a range of around +/- 20% for the $\sigma_i$ parameters from the initial estimate obtained from the BI of the radial distribution functions. We report the final range for all parameters $[\varepsilon_i, \sigma_i]$, ($i = 1, ...,7$) in Supplementary Table 1 of the Supplementary Note 7. Our ranges were post-validated by our final calibration, as discussed in the following.

We trained the Gaussian process surrogate models on 700 simulation samples of the CG DBEGA+PACM system, which also allowed us to fine-tune the extremes for the calibration parameters. Then we trained 500 simulation samples of the CG DGEBA+D400 system, where fewer simulations where needed thanks to the initial fine-tuning. With these surrogates it was possible to perform a variance-based sensitivity analysis, as reported in Figure 3. This type of analysis provided insight into how the responses of the surrogate models depend on their inputs[57,58].

As one would expect, the analysis revealed a strong influence of the $\sigma_i$ parameters on the density, while the dynamics and mechanical properties of the system depend more on the cohesive energies $\varepsilon_i$. This separation was already assumed in previous ER models[40] and it was confirmed here. Since the main sensitivity (white, thinner bars) dominates the total sensitivity (which includes the higher-order interaction effects between the input parameters) in all cases, the response of the CG model can be approximated with a first-degree polynomial. This also suggests that many of the functional relations between the forcefield parameters and DC can be described through a linear function, since the target responses presented in Figure 2 are also close to linear. The relative contribution of the different cohesive energies to our target properties is similar for $\langle u^2 \rangle$, Young's modulus, and yield stress. DC is as relevant as the cohesive energies for $\langle u^2 \rangle$ and yield stress, while its role is suppressed for the Young's modulus. We notice the prominent influence of the parameter $\sigma_6$ on all four measures used here to quantify the mechanical and dynamical properties of the DGEBA+D400 network. This is expected, as bead 6 is a relatively large bead in the repeated unit of the longer D400 molecule. As such, bead 6 makes up for around 28% of all the CG beads of the network, and close to 40% in terms of the bead volume. Variations of $\sigma_6$ lead to large changes in the density of the system, as well as dynamics and mechanical properties.

**CG force field optimization and validation**

Before the calibration of the CG force-field, we needed to identify a flexible candidate class of calibration functions for the nonbonded parameters of the CG model. Previous ER papers[40-42] for simple glass-forming polymers used a sigmoid function for the temperature dependence of cohesive energy and bead size with temperature. The choice is theoretically supported[38] by the transition from the Arrhenius regime of liquids at high temperature to the glassy regime below the glass transition temperature $T_g$, with the supercooled phase in between dominated by the caging dynamics and α-relaxation processes. We initially assumed a similar sigmoidal function for DC, roughly equating an increase in DC to a decrease in temperature given that both actions slow down dynamics. We found this constraint to be too restrictive for our systems: minimizing the discrepancy between the AA and CG response ( equation (3) in our methods section) did not yield a reasonable parametrization using sigmoid functions alone, as shown by Supplementary Figures 4 and 5 in the Supplementary Note 3.

To uncover what functions best describe the DC dependence of the 14 non-bonded parameters, we employed a class of radial basis functions (RBF) described in our methods section. We assumed that each calibration function shares the same shape parameter $\omega$ and that we have three centers for each calibration parameter $\mathbf{x} = [0\%, 50\%, 100\%]$. The number of centers can be increased to capture more complex behavior, but at the cost of overfitting the data and getting unrealistic approximations of the 'true' calibration functions. Our goal was to obtain the simplest force field that is still able to capture the response of the system. To demonstrate the effect of an overfitting parametrization, we include an example in the Supplementary Note 4 (see Supplementary Figures 6 and 7) where the model has been calibrated at DC = 5% increments without analytical description.

The approach described so far using RBF for all the parameters gave us a possible solution for our force-field (see Supplementary Figures 8 and 9 in the Supplementary Note 5), but at the cost of a highly complex parametrization. We wanted to simplify our model by reducing the degrees of freedom of the parametrization without affecting the model's accuracy. Given that our CG and AA models have intrinsic uncertainty that is approximated with our Gaussian process models through the assumption of homoscedasticity, we calculated the probability that for a specific set of calibration parameters the CG models came from the same distribution as the AA models through an objective function that captures the goodness of fit:

$$\mathcal{L}(\boldsymbol{\varepsilon}, \boldsymbol{\sigma}) = \int_0^1 \prod_{i=1}^4 \int_y P\left(f_{i,\mathrm{P}}^{(\mathrm{CG})}(\mathrm{DC}, \boldsymbol{\varepsilon}_\mathrm{P}(\mathrm{DC}), \boldsymbol{\sigma}_\mathrm{P}(\mathrm{DC})) = y\right) P\left(f_{i,\mathrm{P}}^{(\mathrm{AA})}(\mathrm{CD}) = y\right) \mathrm{d}y \mathrm{dCD} +$$

$$\int_0^1 \prod_{i=1}^4 \int_y P\left(f_{i,\mathrm{D}}^{(\mathrm{CG})}(\mathrm{DC}, \boldsymbol{\varepsilon}_\mathrm{D}(\mathrm{DC}), \boldsymbol{\sigma}_\mathrm{D}(\mathrm{DC})) = y\right) P\left(f_{i,\mathrm{D}}^{(\mathrm{AA})}(\mathrm{CD}) = y\right) \mathrm{d}y \mathrm{dCD}, \quad (1)$$

where the subscript corresponds to the $i^{th}$ response variable. Equation (1) has similar properties as a likelihood function and thus lends itself to be used in an approximate Bayesian computation scheme to get a posterior approximation of the parameters that make up the calibration functions. Through a quasi-random sampling scheme, we approximated the first two statistical moments of the calibration functions.

The green curves in Figure 4 show the functions in the RBF class that maximize the objective function of the CG and AA models yielding the same target properties, where the uncertainty quantification for each function is also reported (green band). Note that some of the calibration functions have a large envelope of uncertainty (e.g., $\varepsilon_6$ and $\varepsilon_7$), while others have a small uncertainty envelope (e.g. $\sigma_2$ and $\sigma_6$). If the uncertainty envelope is small, we were able to make a well-informed decision on the class of functions that would be most suited to model the non-bonded force field relation to DC. When the uncertainty bounds are large, then the choice of function is not consequential to the calibration accuracy, and we were able to simplify the function. In essence, the quantified uncertainty provides a decision support tool that gives modelers insight into what calibration functions are most significant to the calibration accuracy. The functions' uncertainty reported in Figure 4 is a local measure of uncertainty around the function mean value considering all the target properties, while the sensitivity analysis of Figure 3 is a global measure in the whole parameter space for each property separately. Still, it is possible to connect the two quantities considering the joint probability distributions. We discuss this briefly in our Supplementary Note 8 (see Supplementary Figures 11 and 12), and we will report these technical findings in detail in an upcoming paper focused on the statistical analysis approach to functional calibration.

With this procedure, it was possible to drastically simplify our parametrization, reducing most functional forms either to linear functions or constants with changing DC. For the simplification, we used the results presented in Figure 4 and considered either a constant function or a linear function if it would fit within the envelope of uncertainty (where we preferred constant over linear as it requires one fewer parameter). With this initial guess, we used Equation (3) (see our methods section) to minimize the squared difference for the new set of calibration functions. The results of this simplification are the black lines in Figure 4: only the parameter $\varepsilon_3$ required an RBF; $\varepsilon_2$, $\varepsilon_5$, $\sigma_1$ and $\sigma_3$ required a linear dependence on DC, while the remaining 9 parameters could be kept constant. The number of free parameters needed for this parametrization was reduced from 43 (all RBF) to 21 (simplified formulation), see Table 2. We note that once an inference has been made on the new class of function that can be used for each parameter in the simplified formulation, the goal is to globally minimize the discrepancy between the AA and CG models response. As such, each simplified function (black curves in Figure 4) is not necessarily an analytical approximation of their respective RBF (green curves). Some of the trends obtained are in line with our expectations, like a general increase of $\varepsilon_3$ with increasing DC as the main parameter to control the system's response, given its preeminent role in determining the dynamics and mechanical properties of the CG model, as observed in the sensitivity analysis shown in Figure 3. The parameters associated with beads 1-3 (the DGEBA molecule) showed the strongest trends. This makes sense, as DGEBA is present in both networks. For the bead sizes in particular, the DC-dependence of both systems is controlled uniquely through $\sigma_1$ and $\sigma_3$, all other bead sizes being kept constant. The increase of $\varepsilon_2$ and $\varepsilon_3$ with increasing DC controls the increase of Young's modulus, yield stress and $\langle u^2 \rangle$ in the DGEBA+D400 network, since $\varepsilon_6$ and $\varepsilon_7$ (part of the D400 molecule) are kept constant. A downward trend of $\varepsilon_5$ (bead of the PACM and D400 molecules) likely compensate the effect of $\varepsilon_2$ and $\varepsilon_3$. We want to stress that this solution might not be unique, within small variations of overall accuracy, and the specific details of these functional calibration parameters will depend on the search space of the algorithm, the details of the training data set and other protocol dependent parameters. This is particularly true for parameters with a large uncertainty envelope, where the model's outputs are not strongly affected by variations of the parameter. But the convergence of the algorithm ensures an excellent match between the target properties in the AA and CG force fields, as we show in the following, which is robust against these variations. For reproducibility purposes, we include in our supplementary materials our complete data set, inputs and outputs of all AA and CG simulations, as well as the LAMMPS input files and structure used to obtain these results.

We report in Table 2 the analytical description of all the parameters in the simplified formulation shown in the black curves of Figure 4. For each parametrization, the ML algorithm predicted the response of the CG model for all target properties as a function of DC, which was compared to the values of the same properties in the AA Gaussian process model through Equation (3). For the parametrization shown in Figure 4, the ML-predicted response of the CG model compared to the AA values is reported in Figure 5. For each target property, the ML extrapolation assigned a confidence interval in addition to the expected value for both the AA and the CG systems, with larger intervals for complex properties like the Young's modulus, that has a higher measurement uncertainty (see Figure 2c) and, for the CG model, large sensitivity to the variation of the force field parameters. The CG prediction is in line with the AA values for all properties and at any DC.

Our parametrization has a high level of accuracy, and we found a fair agreement[59] (average RMSRE = 10%) between the AA and CG responses. We also note that the limit on the accuracy of our prediction lies in the competition between the different responses (dynamics and mechanical properties in particular), and the ML protocol proposed is able to obtain a much higher accuracy if calibrated on individual responses separately, as shown in Supplementary Figure 10 of our Supplementary NOTE 6. A perfect calibration of $\langle u^2 \rangle$ for the high-DC systems for example (Figure 5a,e) would require a lower mobility of the CG model, which would increase the value of the Young's modulus (Figure 5c,g) above the target AA value. Our optimization provided the best solution taking into account the simultaneous calibration of the targets. Additionally, this protocol is easily generalizable to any system, for any set of target properties. Higher accuracy can be achieved, if needed, at the cost of a more complex force field. We discuss other possible parametrizations in our Supplementary Notes. We note that the framework here developed can be generalized to different systems of high chemical complexity, where a tradeoff between accuracy and generality of the CG force field must be considered depending on the goal and application of the model. Our method can be readily applied to multi-objective parametrizations, where proper weights are attributed, tailoring the force field to specific applications.

Finally, we discuss the results of the CG simulations performed with the parameters reported in Table 2. The stars in Figure 5 correspond to the values of the target properties extracted from CG simulations performed with the simplified parametrization of Figure 4, showing the agreement between the CG Gaussian process prediction and the actual CG simulation.

**CG model predictivity beyond target properties**

With the validated approach and optimized CG force field parameters, we now report the overall dynamics and mechanical response of the CG and AA systems with varying DC.

Figure 6 shows the MSD and stress-strain curves up to 20% tensile deformation for both DGEBA+PACM and DGEBA+D400 systems at DC=0%, 50%, and 90-95% (for PACM and D400 respectively). The CG curves validate the prediction of the ML model and show good agreement with the AA values for $\langle u^2 \rangle$, Young's modulus and yield stress of the systems. In addition to that, the comparison with the AA curves of corresponding DC shows that by matching modulus and yield stress, we captured the overall stress under tensile deformation for the system. By matching the Debye-Waller factor $\langle u^2 \rangle$ we expected to match perfectly the overall MSD curve at longer timescales, given theoretical relationships linking the picosecond caging dynamics to the segmental dynamics of glass-forming systems and validated in previous ER models for simpler homopolymers. For the current model, we do not find a strong evidence of this. Despite matching the picosecond caging dynamics of the AA and CG systems, the AA has faster dynamics at longer timescales for the uncrosslinked systems. We are not sure of the origin of this effect, but it could be caused by the variety of CG beads with different sizes and cohesive energy, which might create a broader spectrum of caging scales and relaxation times. Despite this discrepancy, the effect is greatly reduced in the fully crosslinked network of interest for experimental applications, where the system is frozen in the network conformation and there is no diffusion.

Overall, the current parametrization showed a high level of accuracy and accounted for the variation in the degree of crosslinking of the network. Even if intermediate DC values might be less practical for this specific system, the problem of the ER for CG models is relevant outside of this particular chemistry, and the protocol outlined in this work can be easily generalized. The developed ML model has aspects of great relevance: (i) it provides reliable insight into unknown physics by accounting for the uncertainty in the training data and the response surface approximations, (ii) it is computationally tractable compared to fully Bayesian parametric and non-parametric calibration schemes that are known to struggle with problems with more than ten parameters[33]. The CG simulations of this study run approximately $10^3$ times faster than the AA systems, simulation size being the same. The increased efficiency of our CG model makes it possible to investigate epoxy networks beyond the nanoscale, for instance to examine factors such as heterogeneity or fracture processes that may exhibit scale dependence.

**DISCUSSION**

The development of new epoxy resin composites for next-generation materials requires an understanding of how the macroscopic properties of the system emerge from its molecular structure, with a level of precision hard to achieve in experiments (like tracking the strain and failure of single covalent bonds), and at scales unachievable with AA MD simulations (from tens of nanometers up to the micrometer scale). CG models can address the shortcomings of AA simulations and focus on critical molecular markers like crosslink density, vibrational modes, structural heterogeneities, and localized fracture at larger scales. Still, the creation of CG models for epoxy resins is in its infancy, because of the high chemical complexity of these systems and the presence of crosslinks. In this work, we developed a CG model for epoxy resins using DGEBA as the epoxy, and either PACM or D400 as the curing agent, in stoichiometric ratio. Our choice is based on recent experimental findings[47] showing that a combination of a stiff hardener (PACM) and a more flexible one (Jeffamines) in the same resin leads to a superior mechanical and ballistic response. This is caused by the presence of nanoscale structural and dynamical heterogeneities, which our model will be suited to address.

Our CG model has been shown to match the dynamics and mechanical properties of a higher-resolution AA model, which is consistent with experimental measures[47,49]. In particular, we employed functional calibration to match the density, Debye-Waller factor $\langle u^2 \rangle$, Young's modulus and yield stress at any degree of crosslinking of the network at fixed temperature $T$=300K. This is an extension of our ER CG protocol, which was used in previous publications to match the dynamics and mechanical properties of simpler glass-forming polymer systems by adjusting the non-bonded interactions of the CG model in a $T$-dependent way. Here the external parameter considered is instead the degree of crosslinking DC of the epoxy network. Additionally, the chemical heterogeneity of our epoxy system required the use of multiple different CG beads (7 in this model), leading to 14 adjustable parameters for the non-bonded interactions ($\varepsilon$ and $\sigma$ for each Lennard-Jones potential, with an arithmetic rule of mixing for cross-interactions). We calibrated all our parameters in a DC-dependent way to simultaneously match the four target properties of the AA system (density, $\langle u^2 \rangle$, Young's modulus and yield stress). To find the optimal set of functional calibration parameters in this high-dimensional space, we developed ML tools that use a training set of CG and AA simulations to get computationally efficient surrogates. We leveraged the properties of the surrogate model to quantify the uncertainty of the calibration functions $[\varepsilon_i(\text{DC}), \sigma_i(\text{DC})]$, $(i \in 1, \ldots, 7)$ for which we initially assumed a relatively high dimension and flexible class of radial basis functions. Subsequently, we used the insight of the uncertainty quantification to greatly simplify the complexity of the calibration functions while maintaining an excellent match between the AA and CG model simulations.

The CG model here reported is $\approx 10^3$ times faster than AA simulations and it will allow the investigation of a broad class of epoxy resins beyond the nanoscale, providing quantitative predictions to explain experimental findings and to guide the design of new materials. By introducing bond-breaking events at large deformations, it would be possible to use this model to study the fracture and impact resistance of epoxy resin networks. Our preliminary results show that this model is robust when multiple curing agents in varying stoichiometric ratio are used, but a more quantitative analysis will be the focus of a future study. Thanks to the larger scales achievable by this model, it will also be possible to investigate the properties of composite systems by adding nanofillers, polymer matrixes or other elements to the resin, at size scales of hundreds of nanometers. The ML tools developed for the parametrization of our model allowed the extension of the energy renormalization CG protocol to a highly complex system with multiple target macroscopic properties. The same scheme can be adopted by the modeling community for the creation of chemistry-specific CG models of arbitrary complexity, coupling physical intuition with the computational power of Gaussian processes for the exploration of the force field parameters space.

**METHODS**

**Systems preparation**

Our simulations were performed with the LAMMPS software[60]. We simulated all-atom systems of either Bisphenol A diglycidyl (DGEBA) and 4,4-Diaminodicyclohexylmethane (PACM) or DGEBA and polyoxypropylenediamine (Jeffamine D-400) in stochiometric ratio for the formation of the cured epoxy resin. For the first system, we placed 768 DGEBA and 384 PACM molecules randomly in a cubic box with periodic boundary conditions. For the second system, we used 944 DGEBA and 472 D400 molecules. We prepared crosslinked networks at intervals of 5% DC, from 0% to 90% (DGEBA+PACM) or from 0% to 95% (DGEBA+D400), DC=0% being the uncrosslinked systems and DC=100% being the fully cured network. We could not achieve higher DC values for the AA networks within reasonable times. We employed the DREIDING force field[61], which we validated for similar epoxy systems in our previous paper[50], showing that the AA model captures the experimental glass-transition temperature and fracture behavior of the fully cured epoxies, and that is compatible with the ReaxFF force field[62] under tensile deformations. We used LAMMPS harmonic style for bond and angles, charmm style for dihedrals, umbrella style for improper interactions and the buck/coul/long pair style for non-bonded interactions. The atomistic molecules were pre-built with no hydrogen atoms in the PACM/D400 amine group and an open-ring configuration for the DGEBA epoxide group, consistent with the final structure after crosslinking. In our previous work, we found that the presence of partial charges on the terminal epoxide and amine groups of uncrosslinked molecules did not have an observable influence on the dynamics and mechanical properties of the system[50]. For each of our systems, we run two independent replicas to enhance the statistics.

For the CG model, we prepared systems of 2000 DGEBA and 1000 PACM molecules, or 1000 DGEBA and 500 D400 molecules. In our CG representation, shown in Figure 1a, we used five beads to represent DGEBA (with only three different bead types due to the molecular symmetry), four beads to represent PACM (of two different types), and fifteen beads (of three different types) to represent D400. This choice allowed us to have independent beads to conveniently use for crosslinking (one for the epoxide group and one for the amino group). The centers of the beads locate at the center of mass of the grouped atoms. We note that other mappings might also work, and have been used in the literature[27]. We think that the capability of our ML protocol is robust to variations in the mapping choice, though rigorous testing of this idea is beyond the scope of this paper. We refer to the DGEBA beads as beads 1, 2, 3; PACM beads as beads 4, 5; D400 beads as beads 5, 6, 7. Bead 5, present both in PACM and D400, corresponds to the amino group $NH_2$ involved in the crosslinking with the epoxide group in DGEBA (bead 3 in the CG representation). We used LAMMPS harmonic style for bond and angles and the lj/gromacs pair style for non-bonded interactions with the arithmetic rule of mixing: $\varepsilon_{ij} = \varepsilon_i \varepsilon_j$ and $\sigma_{ij} = (\sigma_i + \sigma_j)/2$, where $\varepsilon_i$ and $\sigma_i$ are the cohesive energy and effective size Lennard-Jones parameters of the $i^{\text{th}}$ bead. The parameters used were extrapolated from the AA simulations: bonded interactions via Boltzmann Inversion[54] and non-bonded interactions via the energy renormalization-informed ML algorithm, as described in our results section.

**Crosslinking protocol**

We used the Polymatic package[63] to create crosslinks in our systems in cycles of polymerization. In each cycle, the Polymatic algorithm created a certain number of new bonds between target beads within a distance criterion, and for each new bond, it updated the topology of the system and performed an energy minimization using LAMMPS. At the end of each cycle, a molecular dynamics (MD) step is performed to further relax the system. The procedure stopped when the desired number of new crosslinks had been created.

For the AA systems, we created bonds between the carbon atoms of the DGEBA epoxide group and the nitrogen atoms of the PACM or D400 amine group within a cutoff distance of 6.0 Å and creating 16 bonds per cycle. The intermediate molecular dynamics step was performed with a timestep of 1 fs for 50 ps in total, in NPT ensemble (constant number of particles, pressure and temperature) at temperature $T=600$ K and pressure $P=1$ atm. In the CG model, we created 10 bonds per cycle between bead 3 of DGEBA and bead 5 of PACM or D400 within a cutoff distance of 15 Å. The intermediate dynamics step has a timestep of 4 ps, runs for 200 ps in total and it is done in NPT ensemble at $T=1000$ K and $P=0$ atm. The CG interactions used for the network creation are the preliminary results obtained via BI, see Figure 2 for details.

Each amine group can be connected to two DGEBA epoxide groups. In the formation of our networks, we first prioritized the crosslinking between an epoxide group and an amine group with no other crosslinks, creating networks with a DC of up to 50%. After that, we created crosslinks between amine groups and epoxide groups of DGEBA molecules that are not already in the same network, to avoid the formation of closed loops involving only a fraction of the molecules of the system. This restriction allows up to 75% crosslinked networks, at which point all molecules of the system are connected to the same network. We applied no restriction after that, and stopped the procedure when the formation of a new crosslink is not achieved within 30 MD cycles. This limit was at DC=90% for the atomistic DGEBA+PACM system, at DC=95% for the AA DGEBA+D400, and at DC>99% for the CG systems. The data production of this work used these networks with varying chemistry and DC as starting points.

**Data production**

After a short run with a non-bonded soft potential at $T$=300 K and $P$=0 atm to remove overlapping atoms, we followed previous annealing protocols[64] to reach an equilibrated state (signaled by zero residual stress in the system) at room temperature and pressure in the NPT ensemble. For the AA systems, we used a timestep of 1 fs. We first increased the temperature to $T$=600 K and the pressure to $P$=1000 atm in 50 ps in NPT ensemble, then equilibrated the system for 100 ps at high $T$ and $P$, then quenched down to $T$=300 K and $P$=0 atm in 100 ps and finally equilibrated at $T$=300 K and $P$=0 atm for 200 ps. The mean square displacement of the systems was calculated after the equilibration, for the following 100 ps, then a tensile deformation was performed in the NPT ensemble at strain rate $\dot{\varepsilon} = 0.5e^9$ s$^{-1}$. $\langle u^2 \rangle$ was calculated from the mean square displacement at $t^* = 3$ ps, following previous protocols[40]. The choice of the timescales was made to obtain an equilibrated system within a reasonable computational time. The tensile deformation was performed separately in three different directions, i.e., x, y and z to obtain improved statistics of the mechanical properties of the systems. The Young's modulus was calculated from the slope of the stress curve during the tensile test within total strain=2%. The yield stress was calculated at the intersection of the stress curve with a fit of the Young's modulus shifted to start at strain=3%.

The CG systems used a timestep of 4 fs. They were first equilibrated at $T$=800 K and $P$=100 atm, then quenched to 500 K and 0 atm to relax the pressure, then quenched in temperature to 300 K and 0 atm, and finally equilibrated at constant $T$=300 K and $P$=0 atm. Each of these simulation phases run for 2 ns. The dynamics was then measured in the equilibrated state to extract $\langle u^2 \rangle$ and density. A tensile test with strain rate $\dot{\varepsilon} = 0.5 \times 10^9 \text{ s}^{-1}$ (same as the AA simulations) was performed in the NPT ensemble to extract the Young's modulus and the yield stress.

**Machine Learning and Functional Calibration**

A key component of the proposed framework is the adoption of Gaussian process ML models to replace our costly AA and CG models and simulations. The motivation for choosing Gaussian process models over other ML models (e.g., in comparison to artificial neural networks[65] and random forests[66]) is that they are data efficient and enable the quantification of prediction uncertainty. The uncertainty quantification allows us to start with a high-dimensional parametrization with many free parameters, and simplifying the final solution based on the predicted uncertainty, as we show in Figure 4. Gaussian processes naturally incorporate the multi-response calibration that we need. Finally, we remark that the convergence of alternative methods such as a particle swarm optimization (PSO) algorithm would require millions of CG simulations even for a 30-dimensional function[67] with exponential growth, whereas our protocol only needed around 1000 CG simulations to converge for a 43-dimensional problem. For the epoxy model of interest, we trained four Gaussian process models for the DGEBA+PACM and DGEBA+D400 systems (two CG and two AA models).

For the Gaussian process surrogates of the CG models, we designed a set of simulations where each simulation is represented by a point in a 15-dimensional hypercube (7 $\varepsilon_i$ and 7 $\sigma_i$ parameters describing the non-bonded interactions of the seven beads, plus DC). Since our two CG networks do not share the same set of CG beads, we created two experimental designs containing samples $\mathbf{x}_{i,\text{P}}^{(\text{CG})} = \{DC, \varepsilon_1, \sigma_1, \varepsilon_2, \sigma_2, \varepsilon_3, \sigma_3, \varepsilon_4, \sigma_4, \varepsilon_5, \sigma_5\} = \{DC, \boldsymbol{\varepsilon}_\text{P}, \boldsymbol{\sigma}_\text{P}\} \in \mathbb{R}^{11}, (i = 1, \ldots, n_\text{P})$ and $\mathbf{x}_{j,\text{D}}^{(\text{CG})} = \{DC, \varepsilon_1, \sigma_1, \varepsilon_2, \sigma_2, \varepsilon_3, \sigma_3, \varepsilon_5, \sigma_5, \varepsilon_6, \sigma_6, \varepsilon_7, \sigma_7\} = \{DC, \boldsymbol{\varepsilon}_\text{D}, \boldsymbol{\sigma}_\text{D}\} \in \mathbb{R}^{13}, (j = 1, \ldots, n_\text{D})$ for the DGEBA+PACM systems and DGEBA+D400 system, respectively, where $n_\text{P}$ and $n_\text{D}$ are the number of simulations. We then created a design of experiments from a Sobol sequence, a type of fully sequential space-filling design that has excellent space-filling properties for any number of simulations[68]. We obtained two sets of training data $\left\{\mathbf{X}_\text{P}^{(\text{CG})}, \mathbf{Y}_\text{P}^{(\text{CG})}\right\} = \left\{\left\{\mathbf{x}_{1,\text{P}}^{(\text{CG})}, \mathbf{y}_{1,\text{P}}^{(\text{CG})}\right\}, \dots, \left\{\mathbf{x}_{n_\text{P},\text{P}}^{(\text{CG})}, \mathbf{y}_{n_\text{P},\text{P}}^{(\text{CG})}\right\}\right\}^\text{T}$ and $\left\{\mathbf{X}_\text{D}^{(\text{CG})}, \mathbf{Y}_\text{D}^{(\text{CG})}\right\} = \left\{\left\{\mathbf{x}_{1,\text{D}}^{(\text{CG})}, \mathbf{y}_{1,\text{D}}^{(\text{CG})}\right\}, \dots, \left\{\mathbf{x}_{n_\text{D},\text{P}}^{(\text{CG})}, \mathbf{y}_{n_\text{D},\text{P}}^{(\text{CG})}\right\}\right\}^\text{T}$, where $\mathbf{y}_{i,\text{P}}^{(\text{CG})}, (i = 1, \dots, n_\text{P})$ and $\mathbf{y}_{j,\text{D}}^{(\text{CG})}, (j = 1, \dots, n_\text{D})$ are tuples that each contain the four responses of interest (i.e., density, $\langle u^2 \rangle$, Young's modulus and yield stress). Using these samples to train two Gaussian process surrogates provided us with functions that approximate our CG models at unobserved sets of input parameters as $f_\text{P}^{(\text{CG})}(\cdot)|\mathbf{Y}_\text{P}^{(\text{CG})} \sim \mathcal{N}\left(\boldsymbol{\mu}_\text{P}^{(\text{CG})}(\cdot), \mathbf{mse}_\text{P}^{(\text{CG})}(\cdot)\right)$ and $f_\text{D}^{(\text{CG})}(\cdot)|\mathbf{Y}_\text{D}^{(\text{CG})} \sim \mathcal{N}\left(\boldsymbol{\mu}_\text{D}^{(\text{CG})}(\cdot), \mathbf{mse}_\text{D}^{(\text{CG})}(\cdot)\right)$ for the DGEBA+PACM systems and DGEBA+D400 system, respectively. In this formulation, $\mathcal{N}(\cdot)$ is a normal distribution, $\boldsymbol{\mu}_\text{P}^{(\text{CG})}(\cdot)$ and $\boldsymbol{\mu}_\text{D}^{(\text{CG})}(\cdot)$ are the mean predictions for each of the four responses, and $\mathbf{mse}_\text{P}^{(\text{CG})}(\cdot)$ and $\mathbf{mse}_\text{D}^{(\text{CG})}(\cdot)$ are the posterior predictive uncertainties. The $(\cdot)$ symbol stands for all the parameters on which these functions depend. Namely, in our case, $\{DC, \varepsilon_1, \sigma_1, \varepsilon_2, \sigma_2, \varepsilon_3, \sigma_3, \varepsilon_4, \sigma_4, \varepsilon_5, \sigma_5, \varepsilon_6, \sigma_6, \varepsilon_7, \sigma_7\}$.

Adopting a similar approach for the AA models, we trained two Gaussian process surrogates $f_\text{P}^{(\text{AA})}(\cdot)|\mathbf{Y}_\text{P}^{(\text{AA})} \sim \mathcal{N}\left(\boldsymbol{\mu}_\text{P}^{(\text{AA})}(\cdot), \mathbf{mse}_\text{P}^{(\text{AA})}(\cdot)\right)$ and $f_\text{D}^{(\text{AA})}(\cdot)|\mathbf{Y}_\text{D}^{(\text{AA})} \sim \mathcal{N}\left(\boldsymbol{\mu}_\text{D}^{(\text{AA})}(\cdot), \mathbf{mse}_\text{D}^{(\text{AA})}(\cdot)\right)$ on data sets $\left\{\mathbf{X}_\text{P}^{(\text{AA})}, \mathbf{Y}_\text{P}^{(\text{AA})}\right\} = \left\{\left\{\mathbf{x}_{1,\text{P}}^{(\text{AA})}, \mathbf{y}_{1,\text{P}}^{(\text{AA})}\right\}, \dots, \left\{\mathbf{x}_{n_\text{P},\text{P}}^{(\text{AA})}, \mathbf{y}_{n_\text{P},\text{P}}^{(\text{AA})}\right\}\right\}^\text{T}$ and $\left\{\mathbf{X}_\text{D}^{(\text{AA})}, \mathbf{Y}_\text{D}^{(\text{AA})}\right\} = \left\{\left\{\mathbf{x}_{1,\text{D}}^{(\text{AA})}, \mathbf{y}_{1,\text{D}}^{(\text{AA})}\right\}, \dots, \left\{\mathbf{x}_{n_\text{D},\text{D}}^{(\text{AA})}, \mathbf{y}_{n_\text{D},\text{D}}^{(\text{AA})}\right\}\right\}^\text{T}$, respectively. Note that for the surrogates of the AA models the only input is $DC$, (i.e., the experimental design is only one dimensional $\mathbf{x}_{i,\text{P}}^{(\text{AA})} = \{DC\} \in$

$\mathbb{R}, (i = 1, \dots, n_P)$ and $\mathbf{x}_{i,D}^{(AA)} = \{DC\} \in \mathbb{R}, (i = 1, \dots, n_D)$ ) and $\mathbf{y}_{i,P}^{(AA)}, (i = 1, \dots, n_P)$ and $\mathbf{y}_{j,D}^{(AA)}, (j = 1, \dots, n_D)$ are tuples that each contain our four responses of interest. Finally, $n_P$ and $n_D$ are the number of simulations for the DGEBA+PACM systems and DGEBA+D400 system, respectively.

A common approach for calibration is to minimize the discrepancy between the CG and the AA model predicted through the surrogate models as

$$\boldsymbol{\varepsilon}^*, \boldsymbol{\sigma}^* = \underset{\boldsymbol{\varepsilon} \in E, \ \boldsymbol{\sigma} \in \Sigma}{\operatorname{argmin}} \sum_{i=1}^{4} \left\| \mu_{i,P}^{(AA)}(DC) - \mu_{i,P}^{(CG)}(DC, \boldsymbol{\varepsilon}_P, \boldsymbol{\sigma}_P) \right\|_{L_2} + \left\| \mu_{i,D}^{(AA)}(DC) - \mu_{i,D}^{(CG)}(DC, \boldsymbol{\varepsilon}_D, \boldsymbol{\sigma}_D) \right\|_{L_2}, \quad (2)$$

where $\|\cdot\|_{L_2}$ is the $L_2$ norm and the subscript corresponds to the $i^{\text{th}}$ response variable. This is a parametric approach that allows the identification of a set of parameters that are constant over the space of $DC \in [0\%, 100\%]$. However, this assumption greatly limits the flexibility of the CG models' responses (i.e., poor calibration performance) We showed in Supplementary Figures 2 and 3 that DC-independent parameters are not sufficient to obtain a good match between the AA and CG models. Consequently, we required that each parameter has a dependence on crosslinking density $[\varepsilon_i(DC), \sigma_i(DC)]$ described analytically from DC=0% to DC=100%. Using the functional representation and by replacing the $L_2$ norm with the sample average taken over $n$ samples gives

$$\hat{\boldsymbol{\varepsilon}}(\cdot), \hat{\boldsymbol{\sigma}}(\cdot) = \underset{\boldsymbol{\varepsilon}(\cdot), \ \boldsymbol{\sigma}(\cdot)}{\operatorname{argmin}} \frac{1}{n} \sum_{j=1}^{n} \sum_{i=1}^{4} \left( \mu_{i,P}^{(AA)}(DC_j) - \mu_{i,P}^{(CG)}\left(DC_j, \boldsymbol{\varepsilon}_P(DC_j), \boldsymbol{\sigma}_P(DC_j)\right) \right)^2 +$$
$$\left( \mu_{i,D}^{(AA)}(DC_j) - \mu_{i,D}^{(CG)}\left(DC_j, \boldsymbol{\varepsilon}_D(DC_j), \boldsymbol{\sigma}_D(DC_j)\right) \right)^2, \quad (3)$$

where $\boldsymbol{\varepsilon}_P(\cdot), \boldsymbol{\sigma}_P(\cdot)$ is the set of calibration functions associated with the nonbonded potentials of the DGEBA+PACM system, and $\boldsymbol{\varepsilon}_D(\cdot), \boldsymbol{\varepsilon}_D(\cdot)$ is the set of calibration functions for the nonbonded potentials of the DGEBA+D400 system. We chose radial basis functions as the class of functions describing $[\varepsilon_i(\cdot)\, \sigma_i(\cdot)]$, $(i = 1, \ldots, 7)$. The general formulation of the RBFs is given as

$$\varepsilon_i(\cdot) = \mathbf{k}^\mathrm{T}(\cdot)\, \mathbf{K}^{-1} \mathbf{c}, \tag{4}$$

where $\mathbf{k}^\mathrm{T}(\cdot)\mathbf{K}^{-1}$ is a vector of weights for the $n_c$ center points $\mathbf{c} = [c_1, \ldots, c_{n_c}]^\mathrm{T} \in \mathcal{C} \subset \mathbb{R}^{n_c}$. These center points capture the value that the approximated non-bonded energies must meet at $n_c$ discrete values of $\mathbf{z} = [\mathbf{z}_1, \ldots, \mathbf{z}_{n_c}] \in [0\%, 100\%]^{n_c}$. From these values, the $i^\mathrm{th}$ element of $\mathbf{k}(\cdot)$ is obtained as $k_i(\mathrm{DC}) = exp\,(\omega(\mathrm{DC} - \mathbf{z}_i)^2$ and the $ij^\mathrm{th}$ element of $\mathbf{K}$ is obtained as $K_{ij} = exp\,(\omega(\mathbf{z}_i - \mathbf{z}_j)^2$. This leaves the centers $\mathbf{c}$ and the shape parameter $\omega \in [-4, 4]$ to be inferred through Eqn. (2). RBFs are highly flexibles and allow us to increase the number of centers without worrying about the bounds of the space $\mathcal{C}$ over which $\mathbf{c}$ has been defined, as we can set it equal to the bounds used to generate the training data set of the CG models. This is important for two reasons (i) we can ensure that we do extrapolate from our Gaussian process surrogate models as the search space is restricted to a hypercube, and (ii) having the search space defined on a hypercube greatly simplifies the optimization scheme as no constraints need to be enforced.

## DATA AVAILABILITY

The authors confirm that the data supporting the findings of this study are available within the article and the supplementary materials. Supporting materials include LAMMPS input files and starting configuration files for AA and CG epoxy structures at all values of DC, an excel file with all the input values and output responses for the target AA simulations and the CG simulations and a word file with the details needed to replicate the Gaussian models. Resources available at

https://doi.org/10.6084/m9.figshare.c.5543514.v2. Additional data are available from the corresponding author Sinan Keten upon reasonable request.

## CODE AVAILABILITY

Input files for the open source software LAMMPS and all necessary parameters needed to implement the Gaussian process modeling are provided in the figshare repository https://doi.org/10.6084/m9.figshare.c.5543514.v2. Additional details on the code used are available from the corresponding author Sinan Keten upon reasonable request.

## ACKNOWLEDGMENTS

This work is supported by the Center for Hierarchical Materials Design (CHiMaD) that is funded by the National Institute of Standards and Technology (NIST) (award #70NANB19H005), As well as from the Departments of Civil and Mechanical Engineering at Northwestern University and a supercomputing grant from Northwestern University High Performance Computing Center as well as the Department of Defense Supercomputing Resource Center. Z. Meng acknowledge startup funds from Clemson University, SC TRIMH support (P20 GM121342), and support by the NSF and SC EPSCoR Program (NSF Award #OIA-1655740 and SC EPSCoR Grant #21-SA05).

## AUTHOR CONTRIBUTIONS

Zhaoxu Meng and Sinan Keten ideated and supervised the research. Wei Chen ideated and supervised the ML approach to the multi-functional calibration. Zhaoxu Meng created the setup of the MD simulations. Andrea Giuntoli and Nitin K. Hansoge performed the MD simulations and data analysis. Anton van Beek developed the Gaussian process tools and performed the functional calibration. Andrea Giuntoli was the main writer of the manuscript. All authors contributed to the writing.

**COMPETING INTERESTS**

The authors declare that there are no competing interests

**FIGURES AND TABLES**

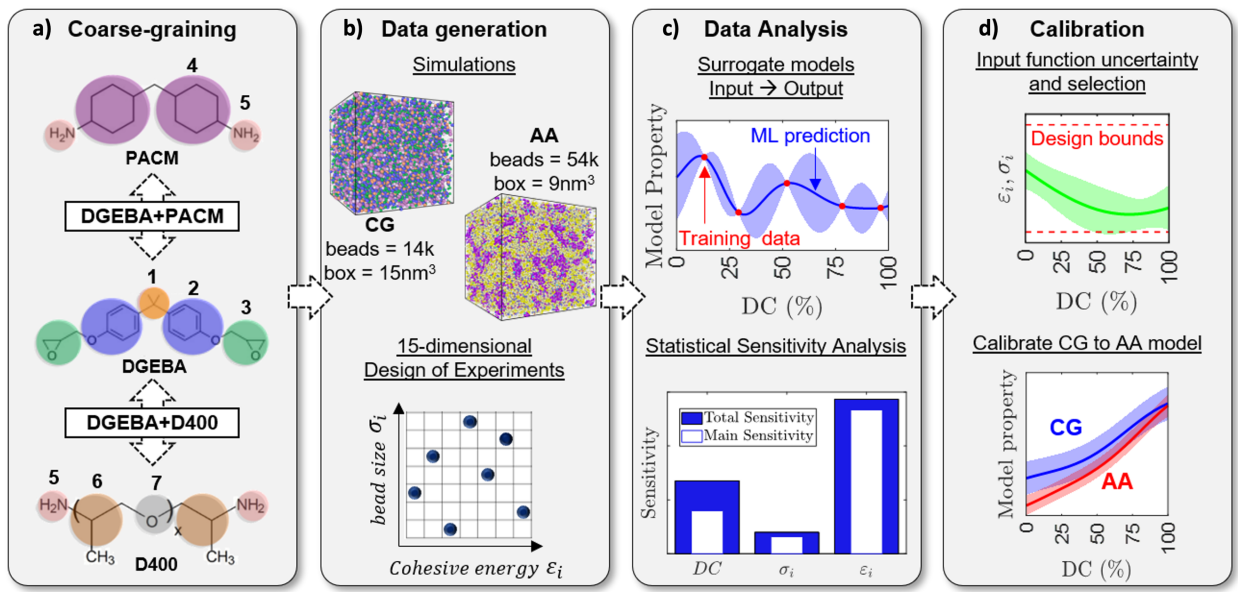

Figure 1: Flowchart of the coarse-graining parametrization protocol. a) Mapping of the CG beads onto the AA chemical structure for DGEBA, PACM, and D400. b) Generation of the training set of CG simulations varying the non-bonded parameters in a 15-dimensional space (two parameters per bead, plus the degree of crosslinking) and generating corresponding system responses. c) Construction of the Gaussian process models from the training set to predict the macroscopic response of the AA simulations for given non-bonded parameters, and sensitivity analysis of each parameter. d) Determination of the optimal values of the CG non-bonded parameters at each DC to match the target properties of the AA models.

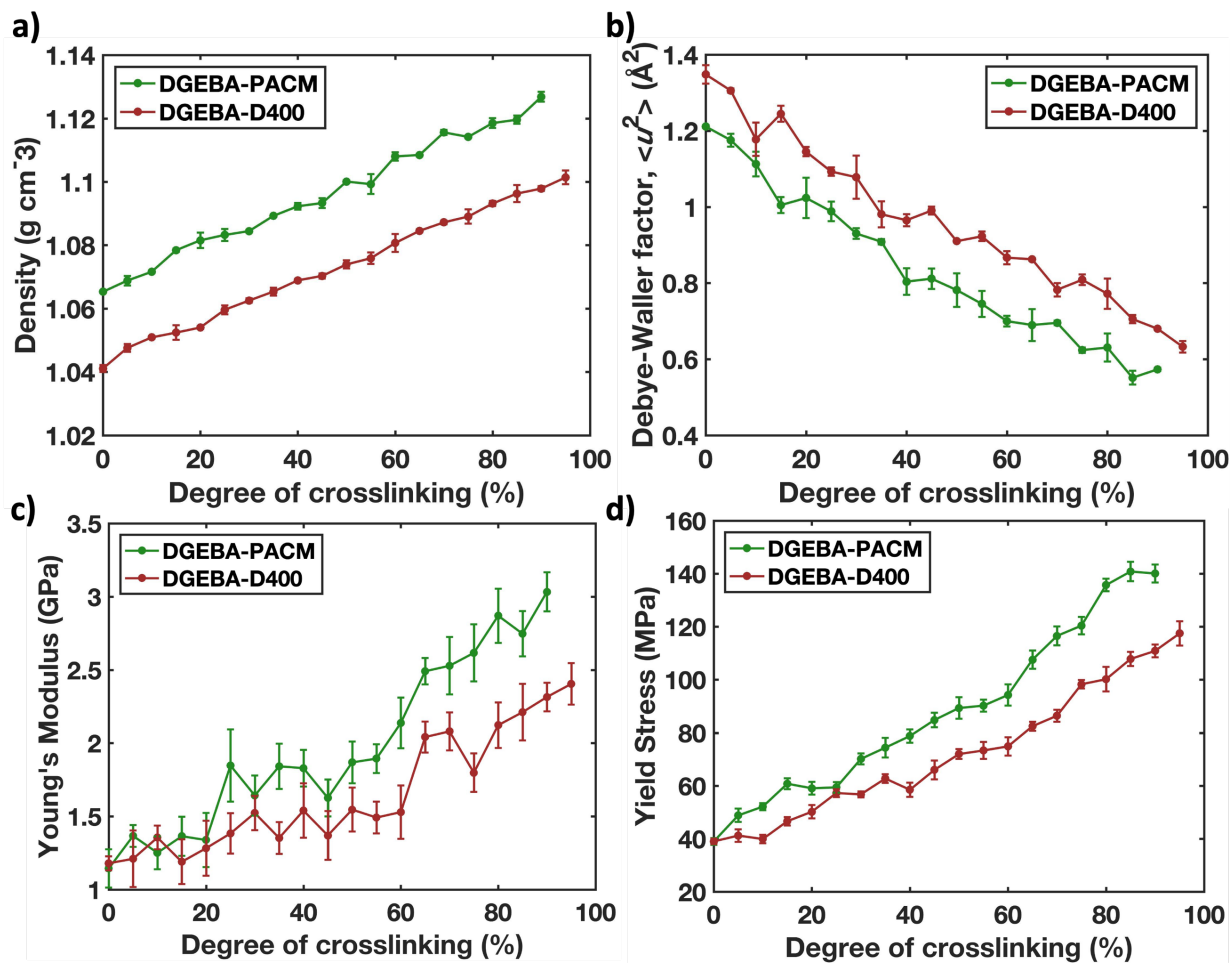

Figure 2: Target macroscopic properties of the AA simulations. (a) density, (b) Debye-Waller factor $\langle u^2 \rangle$, (c) Young's modulus, and (d) yield stress as a function of DC for the DGEBA+PACM and DGEBA+D400 systems. Error bars result from the variance of statistically independent simulations. Density, modulus, and yield stress increase with increasing DC, while $\langle u^2 \rangle$, related to the mobility of the system, decreases. The D400 system, with the longer and flexible curing agent, has a lower density, higher mobility, and softer mechanical response. The dependence of these properties on DC is different in the CG model due to the different changes in configurational entropy caused by the reduction in degrees of freedom. This is typically discussed for changes in temperature, and here observed during the curing process of the polymer network. For this reason, a DC-independent parametrization of the CG model cannot fully capture the features of the AA model at all DC values (see Supplementary Figures 2 and 3), and an energy renormalization procedure is needed.

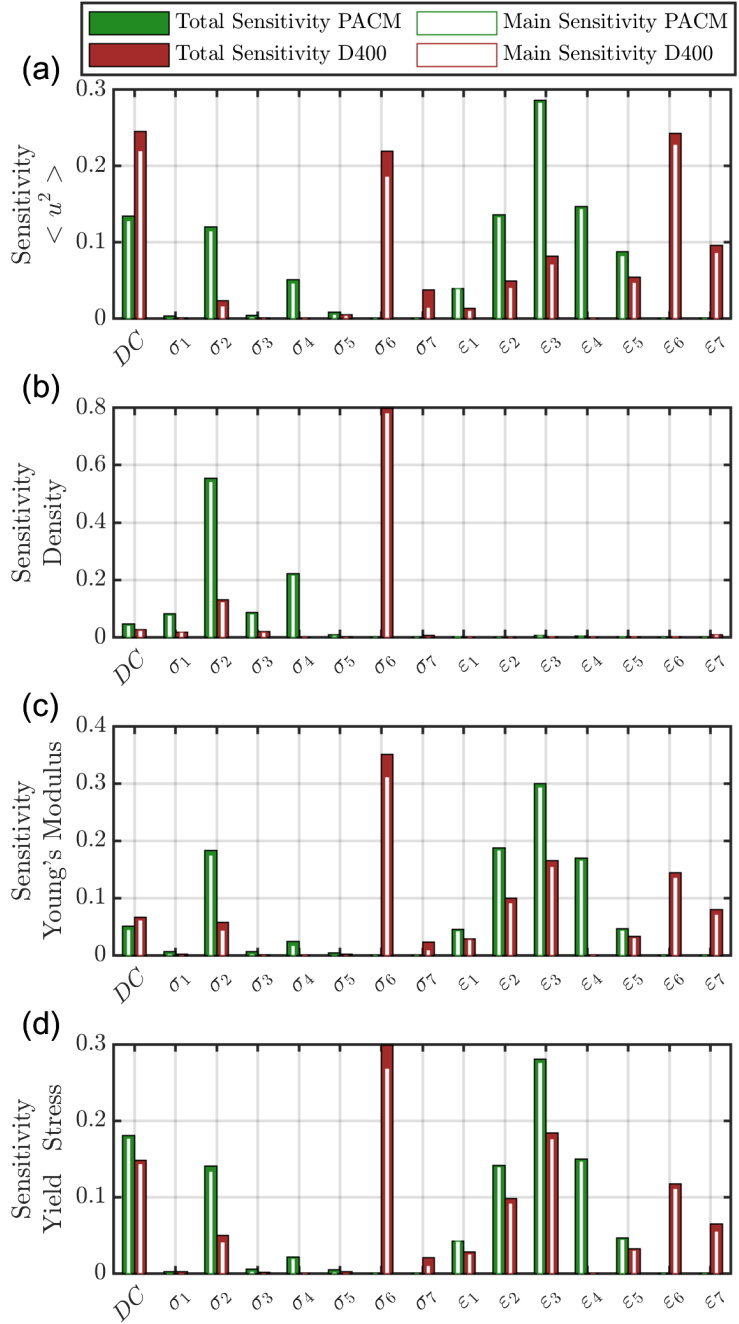

Figure 3: Sensitivity analysis of the target properties varying $[\varepsilon_i, \sigma_i]$ and DC across 1200 CG simulations. The main sensitivity index measures the effect of varying a single input variable on the output. The total sensitivity analysis measures how changing a single input variable affects its contribution to the variance of an output measure while accounting for its interaction with the rest of the input parameters. The density of the systems (panel b) is dominated by the $\sigma_i$ variables, as one would expect. Interestingly, DC has a stronger effect on $\langle u^2 \rangle$ (a) and the yield stress (d) than on the density and the Young's modulus (c). The analysis sheds light on the role of different cohesive energies on the dynamics and mechanical properties of the systems, and it is a useful tool to guide the ML parametrization with the physical insight gained on the model.

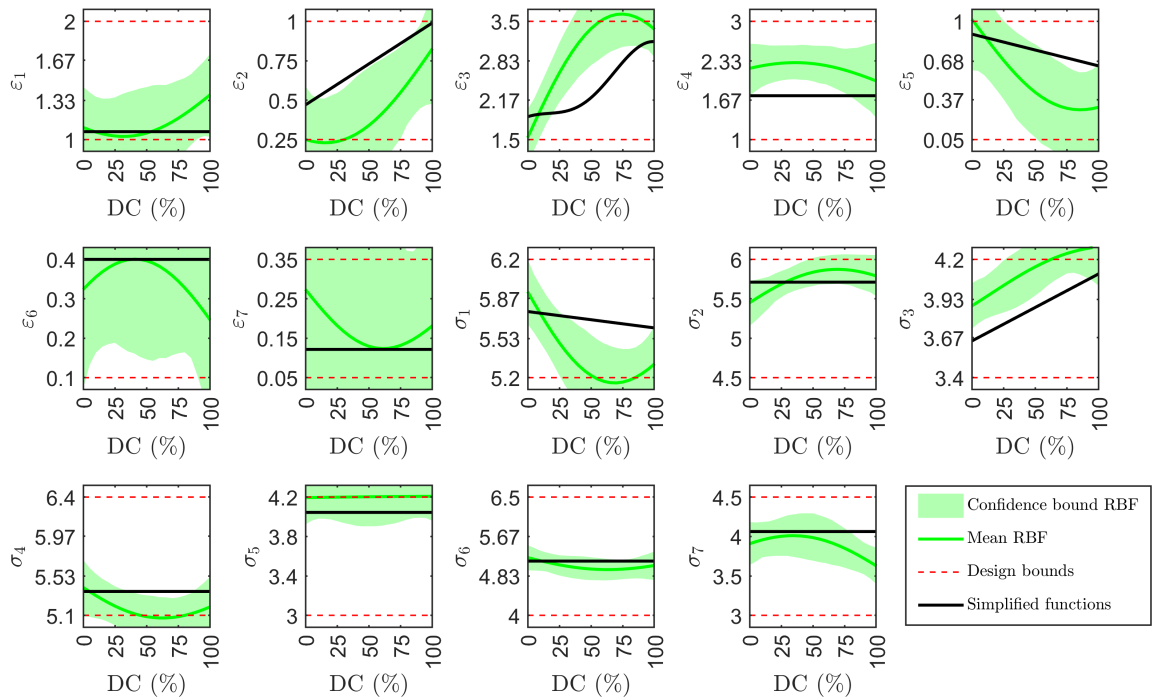

Figure 4: Optimized DC-dependent functions of the non-bonded force field parameters $[\varepsilon_i(DC), \sigma_i(DC)]$. The green curves are RBFs yielding maximum likelihood, see equation (1), between the AA and CG target

properties. The green bands quantify the uncertainty of each parameter, which tells us how sensible the final response of the model depending on the parameter. Where large uncertainties are present, e.g., in the $\varepsilon_6$ and $\varepsilon_7$ functions, we were able to modify the class of function of that parameter to either linear or constant without loss of accuracy of the model's response, thus simplifying the parametrization. The black curves are obtained after simplifying the class of functions and minimizing the squared difference in the AA and CG model response. Note that once a new class of functions is chosen, the new function is not necessarily an approximation of the RBF for each individual parameter. The simplified formulation maintained a fair match[59] with the AA models with an average root mean squared relative error (RMSPE) of 10%. We did not observe a noticeable loss of accuracy of the model compared to calibrations of much higher complexity, see Supplementary Figures 7 and 9.

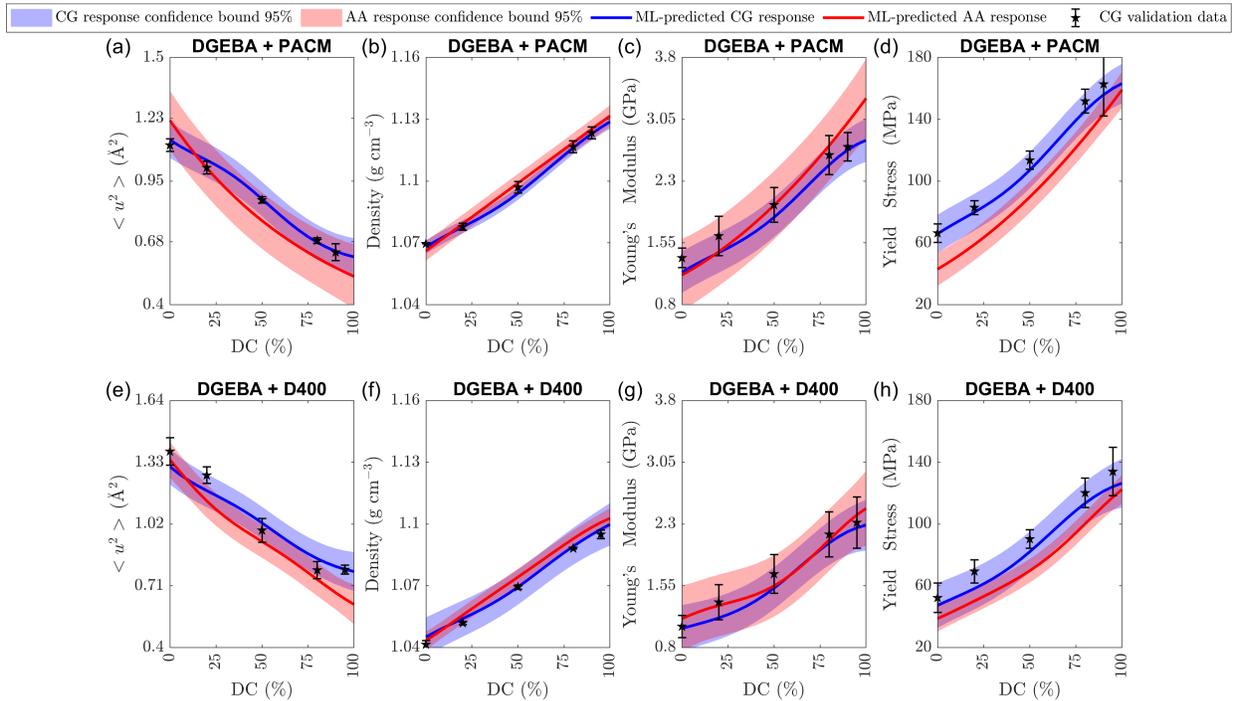

Figure 5: Validation of the predictive power of the Gaussian process extrapolation. Comparison of the target properties as a function of DC between the Gaussian process AA model (red lines), the CG model (blue

lines) with the simplified parametrization shown in Figure 4, and the results of the corresponding CG simulations (black stars). Debye-Waller factor, density, Young's modulus and yield stress are reported for the DGEBA+PACM system (panels a-d) and for the DGEBA+D400 system (panels e-h). The confidence intervals were obtained from the data of Figure 2 for the AA simulations and the design of experiments simulations for the CG model. The error bars on the black stars result from the variance of statistically independent CG simulations. The parametrization of Figure 4 gives a fair agreement[59] for all our targets from the uncrosslinked systems to the fully crosslinked epoxy networks (average RMSPE = 10%). The CG simulation data are in line with the ML-CG prediction, and close to the AA prediction. Slightly higher accuracy is possible with different parametrizations, but at the cost of greatly increasing the complexity of the force field. We discussed other formulations in our Supplementary Notes.

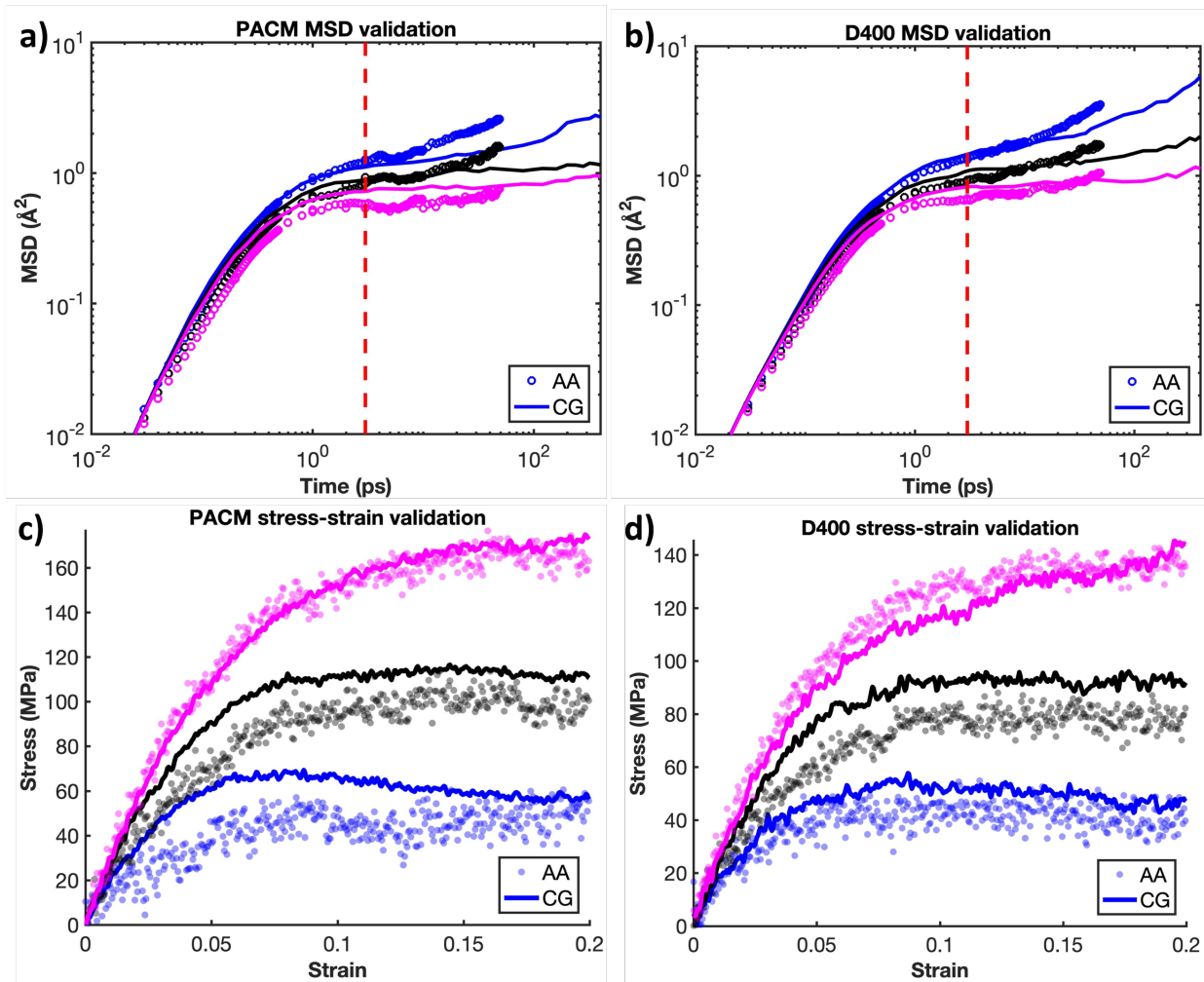

Figure 6: CG validation of the ML parametrization. For our DGEBA+PACM and DGEBA+D400 systems the CG parameters chosen for the non-bonded interactions not only match the target properties we selected (as shown in Figure 5) but can also predict the whole MSD (panels a-b) and tensile stress curves (panels c-d), validating our choice of targets as good predictors of the systems dynamics and mechanical properties.

Table 1: List of all the bonded interaction parameters of the CG model obtained from Boltzmann inversion of the distributions of bonds and angles in the AA simulations, calculated between the centers of mass of the corresponding CG beads.

| Interaction $U_{ij}(l) = k_{ij}(l - l_{ij})^2$ $U_{ijk}(\theta) = k_{ijk}(\theta - \theta_{ijk})^2$ | $k$ kcal/mol·Å² kcal/mol | $l$ or $\theta$ Å degrees |
|---|---|---|
| Bond 1-2 | 201 | 3.37 |
| Bond 2-3 | 22.18 | 4.65 |
| Bond 4-4 | 30.25 | 4.60 |
| Bond 4-5 | 11.87 | 3.32 |
| Bond 5-6 | 49.72 | 1.88 |
| Bond 6-7 | 114.6 | 1.86 |
| Bond 3-5 | 21.48 | 2.58 |
| Angle 1-2-3 | 28.52 | 165 |
| Angle 2-1-2 | 45.60 | 108 |
| Angle 4-4-5 | 7.18 | 160 |
| Angle 7-6-7 | 38.77 | 138 |
| Angle 6-7-6 | 43.62 | 161 |
| Angle 5-6-7 | 38.01 | 138 |
| Angle 2-3-5 | 3.52 | 120 |
| Angle 3-5-4 | 7.49 | 124 |
| Angle 3-5-6 | 9.15 | 117 |
| Angle 3-5-3 | 11.45 | 120 |

Table 2: Parameters for the simplified analytical description of all cohesive energies and bead sizes, as shown in Figure 5. 21 free parameters are needed to describe the 14 non-bonded parametric functions. The analytical expression of the $\varepsilon_3$ RBF function can be found in our methods section.

| Interaction | Functional Form | Interaction | Functional Form |
|---|---|---|---|
| $\varepsilon_1(DC)$ | 1.07 | $\sigma_1(DC)$ | $5.76 - 0.14 \times DC$ |
| $\varepsilon_2(DC)$ | $0.47 + 0.52 \times DC$ | $\sigma_2(DC)$ | 5.71 |
| $\varepsilon_3(DC)$ | $\mathbf{k}^T(\cdot)\mathbf{K}^{-1}[1.89, 2.21\ 3.16,]^T$, $\omega = -0.44$ | $\sigma_3(DC)$ | $3.65 - 0.45 \times DC$ |
| $\varepsilon_4(DC)$ | 1.74 | $\sigma_4(DC)$ | 5.36 |
| $\varepsilon_5(DC)$ | $0.90 - 0.26 \times DC$ | $\sigma_5(DC)$ | 4.04 |
| $\varepsilon_6(DC)$ | 0.40 | $\sigma_6(DC)$ | 5.15 |
| $\varepsilon_7(DC)$ | 0.12 | $\sigma_7(DC)$ | 4.06 |

# Supplementary Information for: Systematic Coarse-graining of Epoxy Resins with Machine Learning-Informed Energy Renormalization

Andrea Giuntoli[1,2], Nitin K. Hansoge[2,3], Anton van Beek[2,3], Zhaoxu Meng[4*], Wei Chen[2,3*], Sinan Keten[1,2,3*]

[1]Dept. of Civil & Environmental Engineering, Northwestern University, 2145 Sheridan Road, Evanston, IL 60208-3109

[2]Center for Hierarchical Materials Design, Northwestern University, 2205 Tech Drive, Evanston, IL 60208-3109

[3]Dept. of Mechanical Engineering, Northwestern University, 2145 Sheridan Road, Evanston, IL 60208-3109

[4]Dept of. Mechanical Engineering, Clemson University, 208 Fluor Daniel EIB, Clemson, SC 29634-0921*To whom correspondence should be addressed. Emails:

zmeng@clemson.edu  (Z. Meng)

weichen@northwestern.edu  (W. Chen)

s-keten@northwestern.edu (S. Keten)

# SUPPLEMENTARY NOTE 1: BONDED POTENTIALS CALIBRATION

In Supplementary Figure 1 we report the AA distributions and CG potentials obtained through iterative Boltzmann inversion of the bonds, angles, and radial distribution functions.

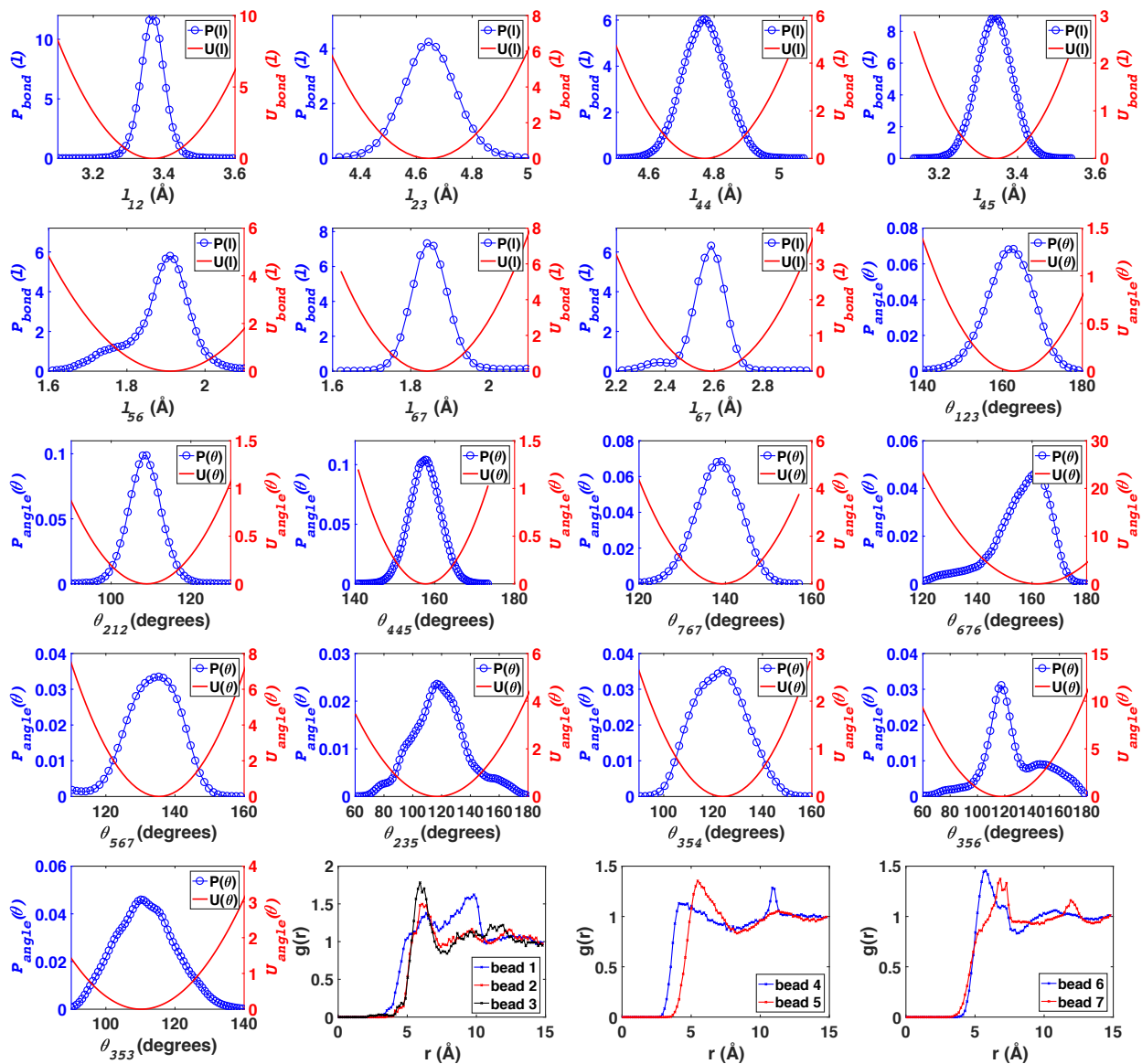

Supplementary Figure 1: Complete set of all distributions of bonds, angles, and radial distribution functions for the centers of mass informed from the AA systems corresponding to the CG beads 1-7 (blue), and the derived CG potentials obtained for bonds and angles using Boltzmann inversion (BI) (red).

The atomistic data are the distributions of the distance/angle between the centers of mass of the corresponding CG beads (calculated as the center of mass of the atoms included in the CG bead), from which the CG bonded potentials are obtained through BI. These bonded potentials are directly adopted in the CG model, which are listed in Table 1 of the manuscript. The last three panels report the radial distribution functions of all seven beads, from which we extract Lennard-Jones (LJ) parameters through BI, which include the cohesive energy ($\varepsilon$) and bead size ($\sigma$) of each type of beads. We then use an arithmetic rule of mixing for cross-interactions between different types of beads. We note that these LJ potential parameters are initial guesses, which are subject to further calibration using the ML approach.

**SUPPLEMENTARY NOTE 2: DC-INDEPENDENT FUNCTIONAL CALIBRATION**

We mention in the main manuscript that a DC-dependence of the non-bonded parameters is necessary to obtain a DC-transferable matching between the AA and CG responses, as the free energy landscape and configurational entropy of the two systems are affected differently by the creation of crosslinks. In Supplementary Figures 2 and 3 we show the model response for a DC-independent calibration which is optimized only for the DC=0% or DC=100% response, respectively. The faster variation of the AA model properties (red lines) with varying DC proves that a DC-independent CG force-field cannot be DC-transferable.

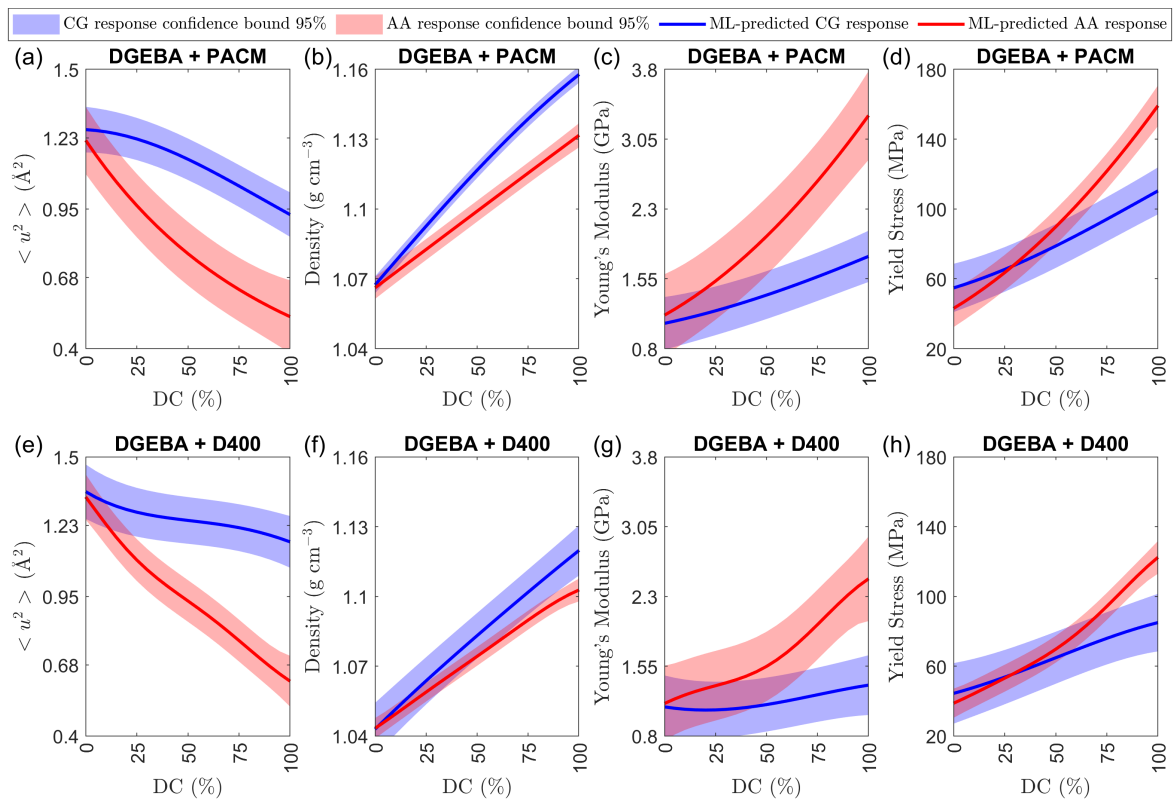

Supplementary Figure 2: Model response for non-bonded parameters optimized for DC=0%. While we obtain an accurate response at DC=0%, the variation of the AA properties with increasing DC is faster than what observed in the CG response.

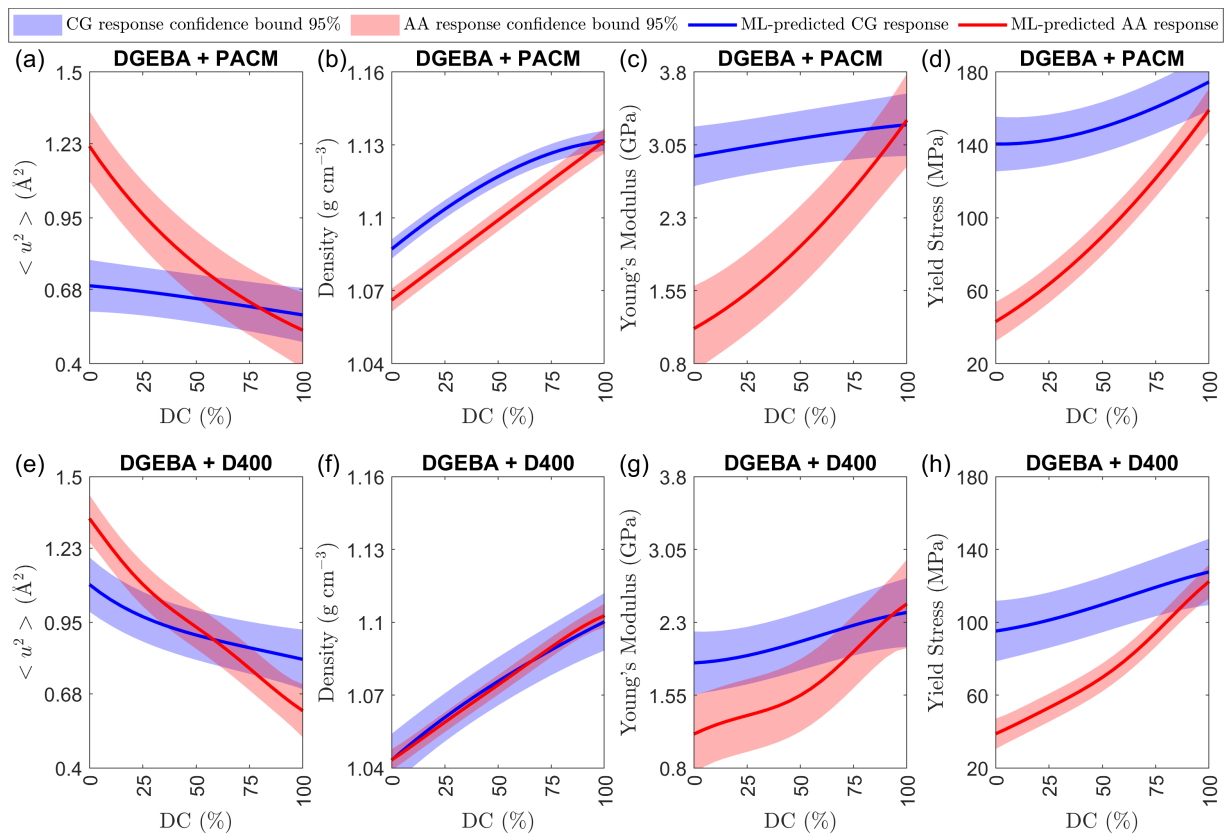

Supplementary Figure 3: Model response for non-bonded parameters optimized for DC=100%. While we obtain an accurate response at DC=100%, the variation of the AA properties with decreasing DC is faster than what's observed in the CG response.

## SUPPLEMENTARY NOTE 3: SIGMOIDAL FUNCTIONAL CALIBRATION

Before resorting to the more complicated radial basis function (RBF) class of functions, we initially attempted a DC-dependent calibration based on sigmoidal functions stemming from the $T$-dependence of the configurational entropy in glass-forming polymers with which we are drawing a parallel. The optimal parametrization using sigmoid functions is shown in Supplementary Figure 4.

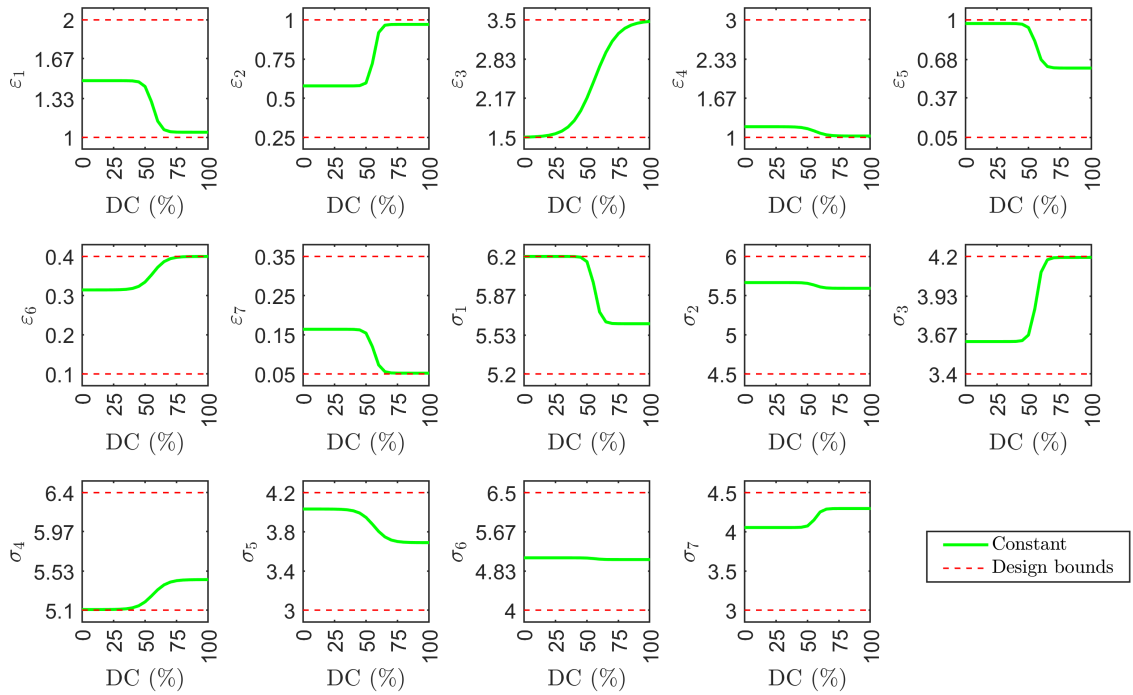

Supplementary Figure 4: Optimal parametrization using sigmoid functions for all parameters. Unfortunately, this solution does not provide an adequate matching between the AA and the CG models, as shown in Supplementary Figure 5. In fact, the same sigmoid-like shape of the parameters is introduced in the CG response, while it is absent in the target AA responses.

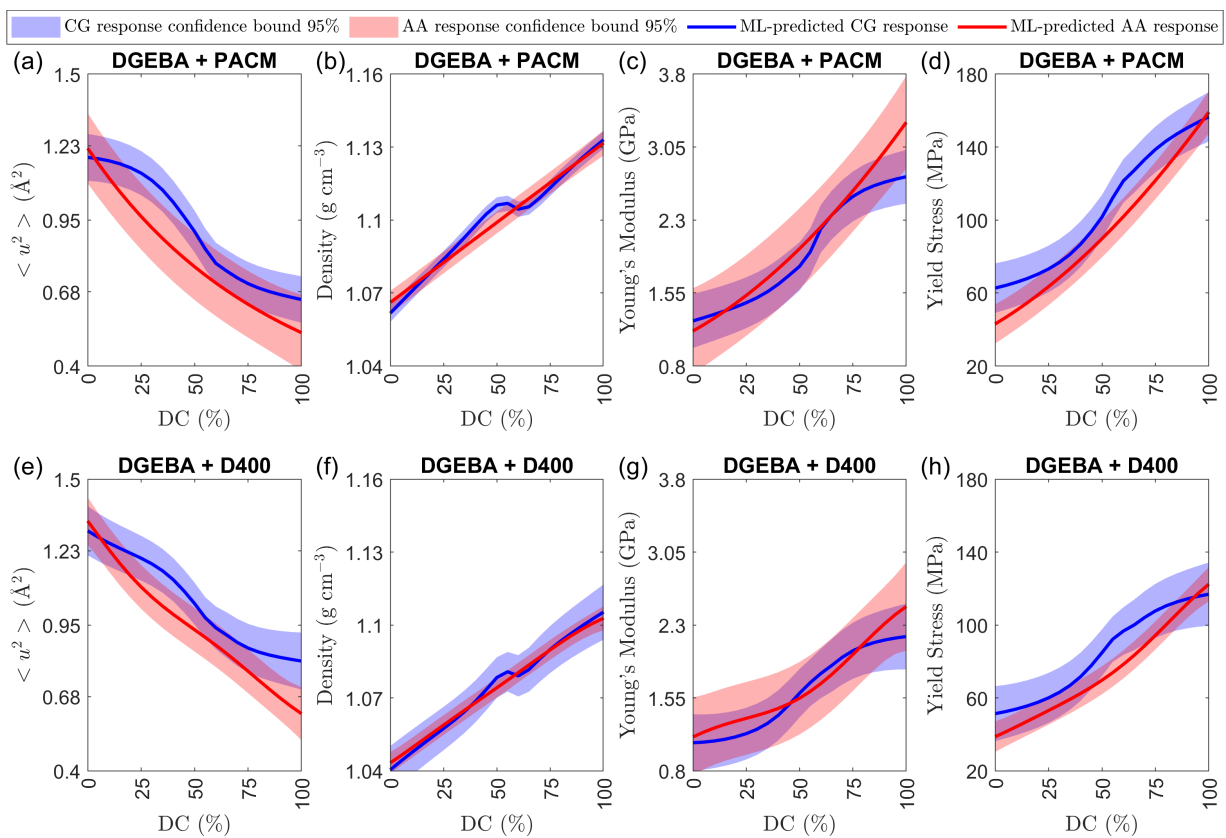

Supplementary Figure 5: Model response with the calibration of Supplementary Figure 4 for the CG model.

## SUPPLEMENTARY NOTE 4: INCREMENTAL CALIBRATION

If one were to abandon an analytical description of the DC-dependence of the non-bonded input parameters, it is possible to obtain a slightly higher level of accuracy (RMSRE = 11%) with a fully numerical description. In Supplementary Figure 6 we show the numerical solution of the optimized parameters performed every 5% in the values of DC. It is clear that the optimal inputs calculated this way are fluctuating wildly, and the same fluctuations are propagated to the model response, see Supplementary Figure 7. As such, despite a numerically higher accuracy, we do not believe that this is a robust approach to the model creation.

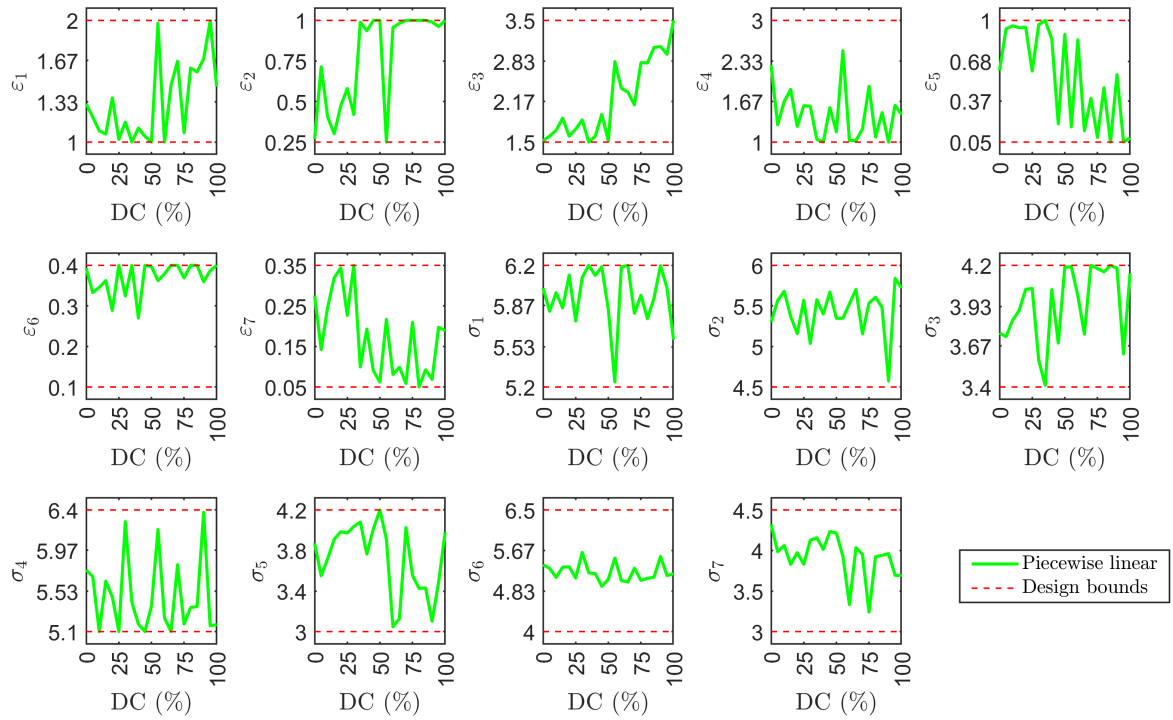

Supplementary Figure 6: Numerical solution for the optimal parameters every DC=5%.

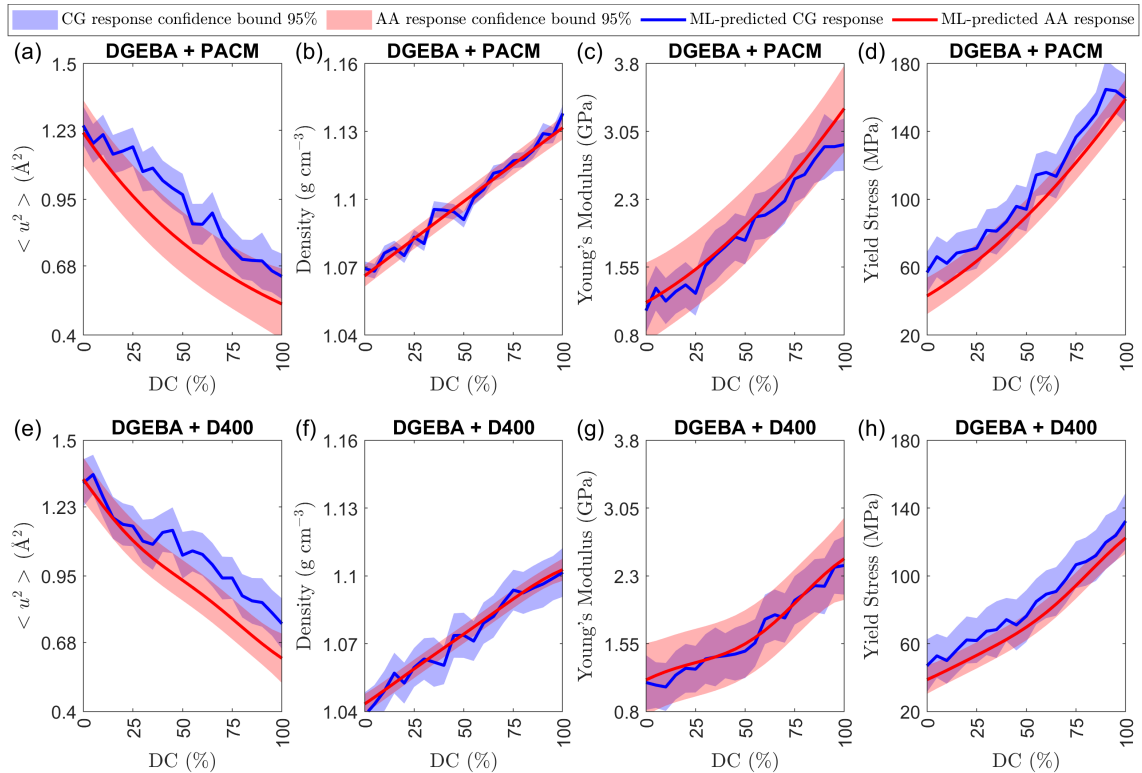

Supplementary Figure 7: Response of the CG model with the numerical solution for the parameters. Despite the slightly higher accuracy (RMSRE=9.6%) the CG model response shows large and discontinuous fluctuations, mimicking the input parameters DC-dependence.

## SUPPLEMENTARY NOTE 5: FULL RBF FUNCTIONAL CALIBRATION

Supplementary Figure 8 reports the calibration obtained using RBF for all our parameters, before the simplification shown by the black curves of Figure 4. Supplementary Figure 9 shows the corresponding response of the model. We note that the accuracy of the response between Supplementary Figure 9 and Figure 5 of the simplified parametrization do not differ substantially, validating our simplification procedure.

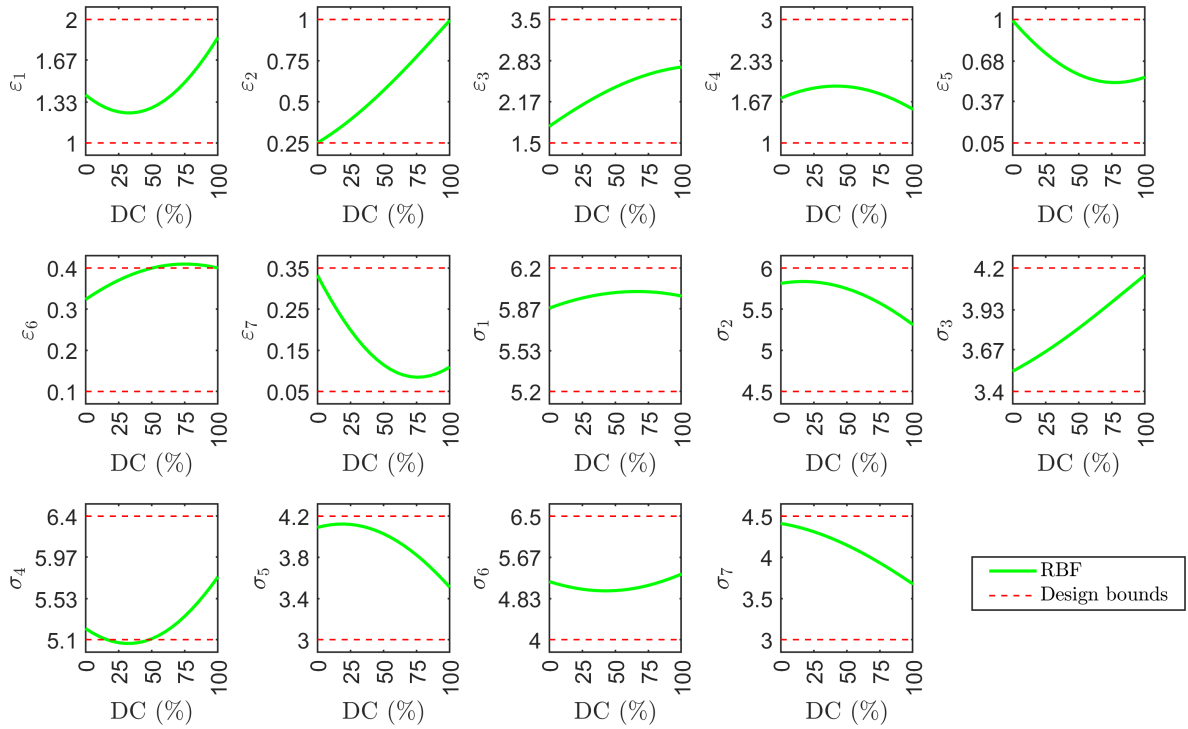

Supplementary Figure 8: Calibration of the variables $[\varepsilon_i(\text{DC}), \sigma_i(\text{DC})]$ as a function of the degree of crosslinking. Each variable is described with an RBF with 3 centers at DC=0%, 50% and 100%, critical values of the polymer network. Within this constraint, the ML algorithm provides the optimal set of curves that minimize the discrepancy between the response of the CG and the AA models at each DC.

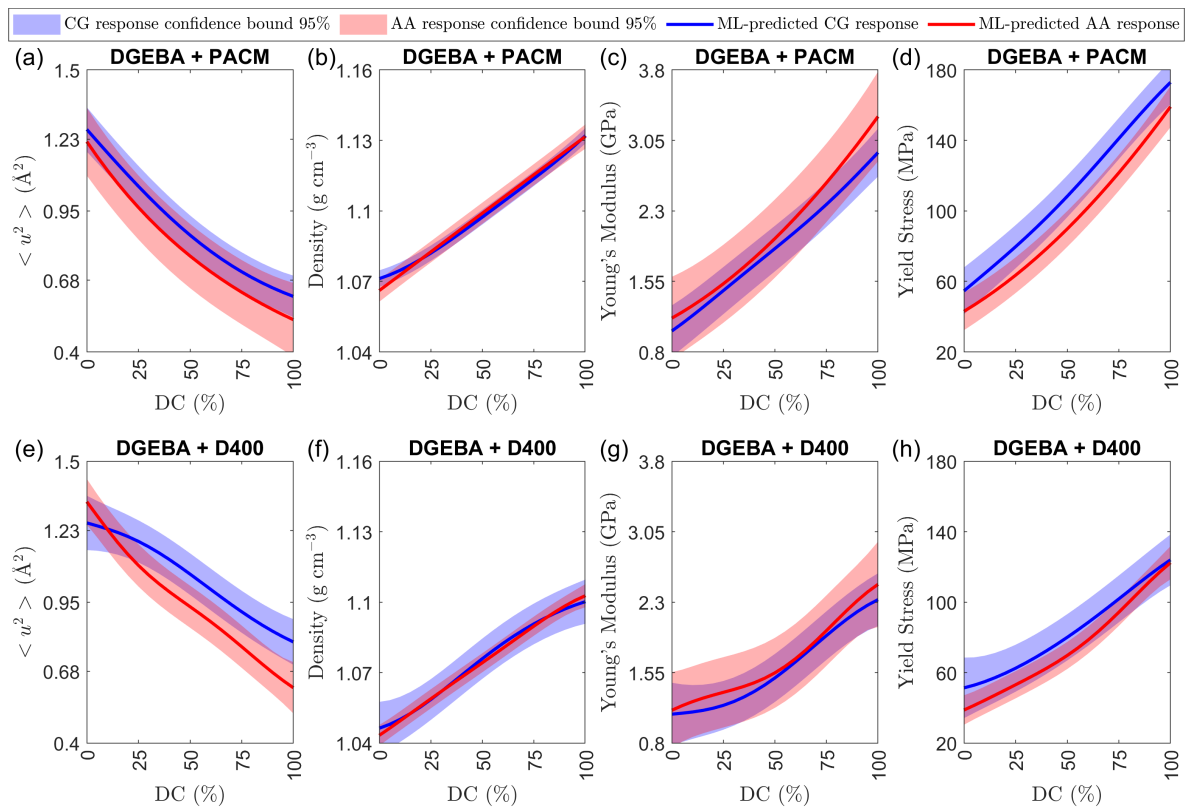

Supplementary Figure 9: Model responses with the calibrated variables $[\varepsilon_i(\text{DC}), \sigma_i(\text{DC})], (i = 1, ..., 7)$ as a function of the degree of crosslinking. The predicted responses are all well in the confidence bounds of the two Gaussian process models: AA (red lines) and CG (blue lines).

# SUPPLEMENTARY NOTE 6: INDIVIDUAL RESPONSE FUNCTIONAL CALIBRATION

The main limit to the accuracy of our final calibration (either using the RBF functions calibration of Supplementary Figure 8 or the simplified parametrization of Figure 4) stems from the requirement of a simultaneous calibration of eight different responses. We show in Supplementary Figure 10 that a calibration aiming to optimize any of the target responses individually can obtain a much higher accuracy, with RMSRE values of around 1% for each response.

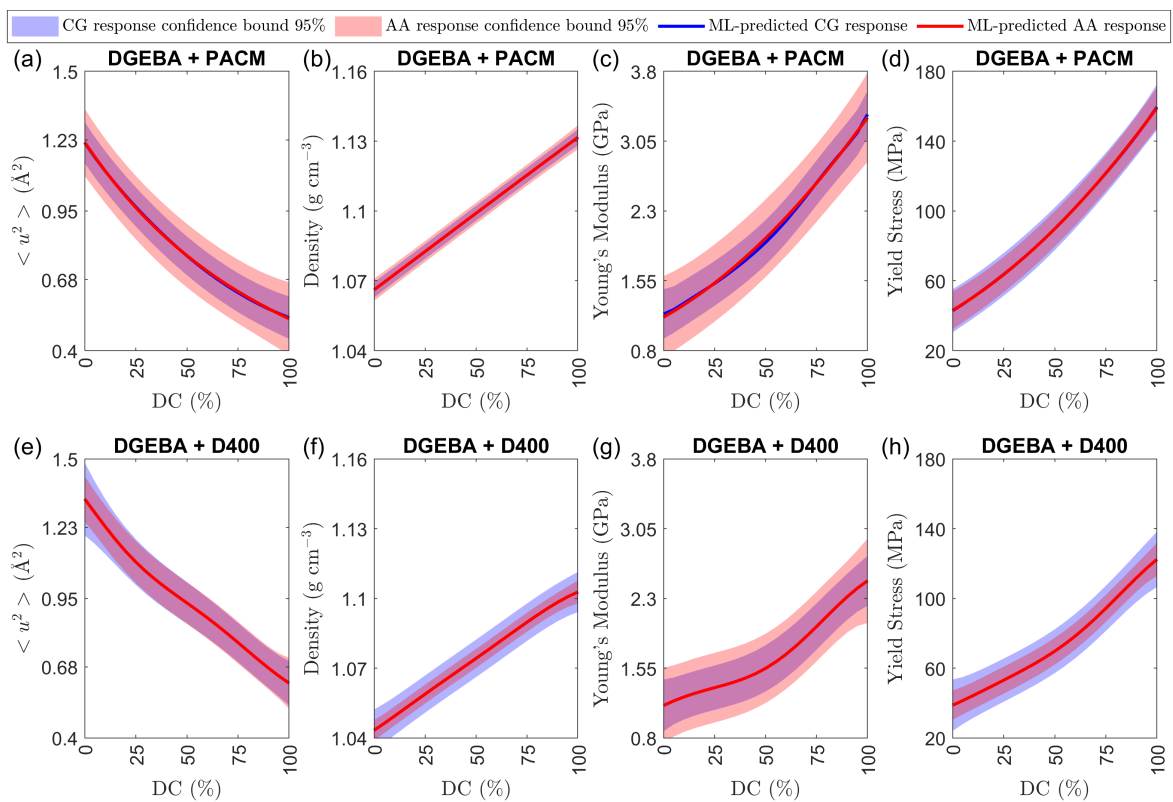

Supplementary Figure 10: CG response for each property when the calibration is performed individually. For each target property it is possible to obtain a high level of accuracy. This shows that the main limit to the accuracy of the simultaneous calibration resides on the competition between physical properties, rather than on the ability of the ML protocol to find optimized parameters.

## SUPPLEMENTARY NOTE 7: THE RANGE OF NON-BONDED PARAMETERS

Supplementary Table 1 reports the range of the non-bonded parameters used for the training set and the calibration of the model, as shown by the dotted lines in Figure 4.

Using the RBF formulation described, the resulting calibration functions are presented in Figure 4, and the corresponding performance predictions are presented in Figure 5. It should be clear that despite the relatively high demand of having to match eight response curves, the RBF functions provide sufficient freedom to minimize the discrepancy between the CG and AA models.

Supplementary Table 1: Ranges of the non-bonded parameters used to train the Gaussian process surrogate models.

| Interaction | Range | Interaction | Range |
| --- | --- | --- | --- |
| $\varepsilon_1$ | $1.0 - 2.0$ | $\sigma_1$ | $5.2 - 6.2$ |
| $\varepsilon_2$ | $0.25 - 1.0$ | $\sigma_2$ | $4.5 - 6.0$ |
| $\varepsilon_3$ | $1.5 - 3.5$ | $\sigma_3$ | $3.4 - 4.2$ |
| $\varepsilon_4$ | $1.0 - 3.0$ | $\sigma_4$ | $5.1 - 6.4$ |
| $\varepsilon_5$ | $0.05 - 1.0$ | $\sigma_5$ | $3.0 - 4.2$ |
| $\varepsilon_6$ | $0.1 - 0.4$ | $\sigma_6$ | $4.0 - 6.5$ |
| $\varepsilon_7$ | $0.05 - 0.35$ | $\sigma_7$ | $3.0 - 4.5$ |

**SUPPLEMENTARY NOTE 8: CORRELATIONS OF PARAMETER FUNCTIONS**

An interesting observation that can be made from the simplified calibration is that most of the simplified functions correspond strongly with the quantified uncertainty as presented in Figure 4. In Supplementary Figure 11, we plot the absolute average (over the three central values of the RBFs) covariance of the calibration functions. These figures elucidate how the various functions are locally correlated. Finally, it is encouraging to observe that $\varepsilon_3$ and $\varepsilon_5$ are strongly correlated with a plurality of other functions, because their associated beads are both a part of the DGEBA+PACM system and the DGEBA+D400 system and correspond to the terminal epoxide and amine groups forming crosslinks.

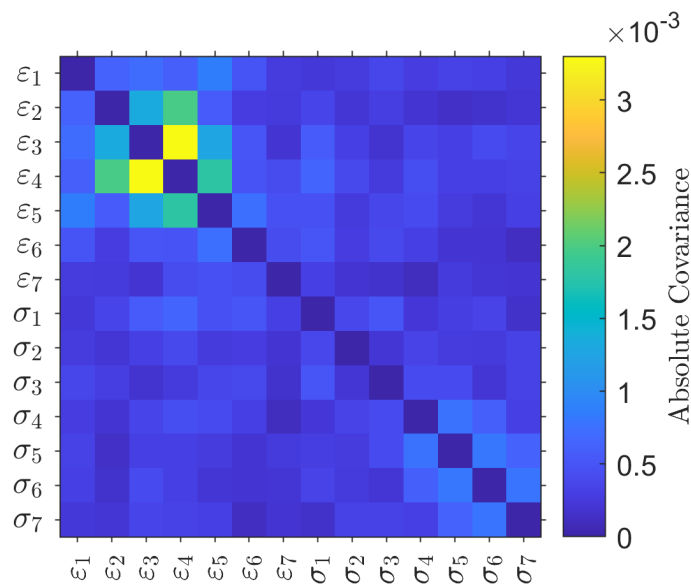

Supplementary Figure 11: Average correlation between the calibration functions of Figure 4 for each calibration parameter. A high correlation between two functions implies that changing one of the two has a significant consequence on the other. We note that $\varepsilon_2, \varepsilon_3, \varepsilon_4$ and $\varepsilon_5$ share a high correlation among themselves and therefore play an important role in the calibration of the CG model. This is an encouraging observation as these functions play a role in both the DGEBA+PACM systems and the DGEBA+D400 system in the formation of the network crosslinks or near the crosslinking sites.

We also notice that it is possible to correlate the global sensitivity analysis shown in Figure 3 to the uncertainty of our RBF functions shown in Figure 4. The sensitivity provides a global description for how the response of the CG simulation depends on the various calibration functions/parameters, whereas the quantified uncertainty depends on how the calibration accuracy changes locally around the set of calibration parameters. In our case, we observed that the CG simulation has a near-linear relation to the calibration parameters (as can be observed from the negligible interaction effects in Figure 3), and as such, the global sensitivity analysis also provides information on the local response. Given this observation, it is reasonable to expect there to be a relationship between the quantified uncertainty and the sensitivity analysis. In fact, in the single response scenario, this relationship is inverse (i.e., a large sensitivity index would correspond to small uncertainty). For visualization purposes, we have plotted the objective function (Equation 1) of individual responses at $DC = 50\%$ to the design range of $\varepsilon_6$ and $\sigma_6$ in Supplementary Figure 12 (eight leftmost panels). We have selected $\varepsilon_6$ and $\sigma_6$ as they correspond to scenarios with small, quantified uncertainty and large quantified uncertainty, respectively. An additional advantage is that they are only inputs to the D400+PACM systems and thus we only need to look at four responses. From the plots in Supplementary Figure 12, we can make two observations:

1) The magnitude of the objective function has a direct relation with the range of the response to simulation noise ratio.
2) If the specific response has a large range to simulation noise ratio then there is an inverse relationship between the quantified uncertainty and the total sensitivity.

The first observation is according to intuition as a relatively large prediction uncertainty in the CG simulation would result in predictive distributions $f_{i,\mathrm{P}}^{(\mathrm{CG})}(\cdot)$ and $f_{i,\mathrm{P}}^{(\mathrm{AA})}(\cdot)$ to have more overlap. However, because these distributions are now wider, their product will be lower and as such the objective function value (Equation 1) will decrease. This can be observed by comparing the modulus with its predicted response. If the range of the response is large to the simulation noise, then we find that the inverse relation between the sensitivity analysis and the quantified uncertainty holds. This can be observed from the sharp peak of $\sigma_6$ in Supplementary Figure 12. Finally, it should be observed that the mode and the width of the individual objective functions greatly influence the shape of the distribution that we obtain when taking their product, as shown by the two rightmost panels of Supplementary Figure 12. Observe that the distribution for $\varepsilon_6$ is much wider than that of $\sigma_6$ and that this agrees with the quantified uncertainty (Figure 4 of the manuscript).

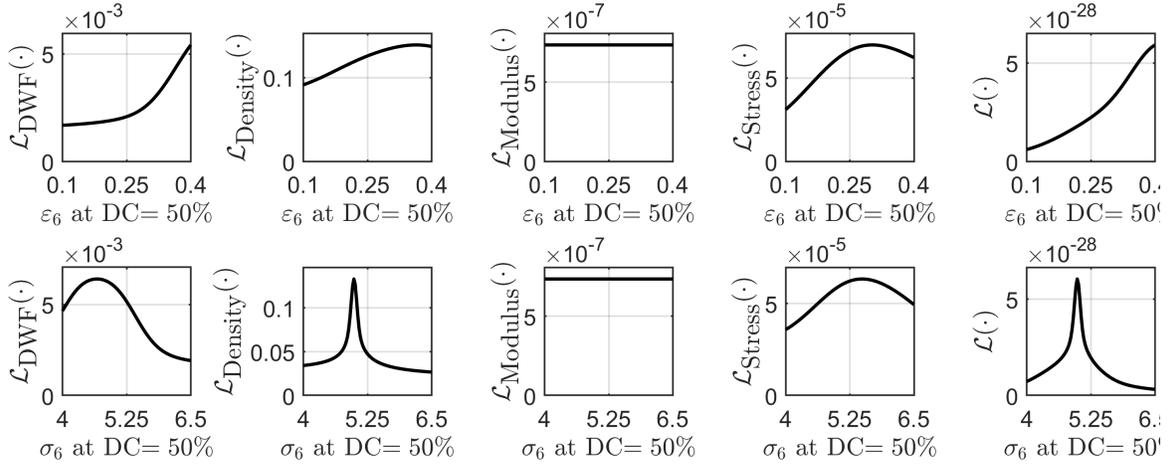

Supplementary Figure 12: Comparison of the objective function value obtained from Equation 1 when we only consider a single response (eight leftmost panels) to the objective function value when we consider all eight responses (two rightmost panels). The objective function values have been plotted for the range of $\varepsilon_6$ and $\sigma_6$. at DC = 50%. The shape of the objective function for all eight responses depends on the width and the mode of the individual distributions.